\documentclass[]{aa}
\usepackage{graphicx}
\usepackage{natbib} 
\bibpunct{(}{)}{;}{a}{}{,}
\usepackage{txfonts}
\usepackage{siunitx}
\usepackage{enumitem}
\usepackage{float}
\usepackage{placeins}
\usepackage{mwe,tikz}
\usepackage[colorlinks=true,linkcolor=blue,citecolor=blue,urlcolor=blue]{hyperref}
 
 \def\Halpha{$\mathrm{H\alpha}$} \def\Hbeta{$\mathrm{H\beta}$} \def\Hgamma{$\mathrm{H\gamma}$} \def\Hdelta{$\mathrm{H\delta}$} 
  \def\HII{\ion{H}{ii}} \def\HeI{\ion{He}{i}}
 \def\HeII{\ion{He}{ii}}  
  \def\CIII{\ion{C}{iii}} \def\CIV{\ion{C}{iv}}
    
  \def\NIV{\ion{N}{iv}} \def\NV{\ion{N}{v}}
   
  \def\OIV{\ion{O}{iv}} \def\OV{\ion{O}{v}}
 \def\OVI{\ion{O}{vi}}

\newcommand{\msunpyr}{\ifmmode{\,M_{\odot}\,\mbox{yr}^{-1}} \else{ M$_{\odot}$/yr}\fi}

\newcommand{\kms}{\ifmmode{\,\mbox{km}\,\mbox{s}^{-1}}\else{km\,s^{-1}}\fi}
\newcommand{\kpc}{\ifmmode {\,\mbox{kpc}} \else{kpc}\fi}
\newcommand{\msun}{\ifmmode M_{\odot} \else M$_{\odot}$\fi}
\newcommand{\rsun}{\ifmmode R_{\odot} \else R$_{\odot}$\fi}
\newcommand{\lsun}{\ifmmode L_{\odot} \else L$_{\odot}$\fi}
\newcommand{\zsun}{\ifmmode Z_{\odot} \else $Z_{\odot}$\fi}
\newcommand{\xsun}{\ifmmode X_{\odot} \else $X_{\odot}$\fi}
\newcommand{\velo}{\ifmmode\varv\else$\varv$\fi}
\newcommand{\vinf}{\ifmmode\velo_\infty\else$\velo_\infty$\fi}
\newcommand{\rgal}{\ifmmode \,R_{\mathrm{gal}} \else R$_{\mathrm{gal}}$\fi}

\begin{document}

 	\title{Spectroscopic and evolutionary analyses of the binary system AzV\,14 outline paths toward the 
 	WR stage at low metallicity}
 	
 	\subtitle{}
 	\titlerunning{Detailed spectroscopic and evolutionary analyses of the massive close binary system AzV\,14.}
 	
 	\author{D. Pauli$^{\ref{inst1}}$ \and L.\,M. Oskinova$^{\ref{inst1}}$ \and W.-R. Hamann$^{\ref{inst1}}$ \and D.\,M. Bowman$^{\ref{inst2}}$ \and H. Todt$^{\ref{inst1}}$ \and T. Shenar$^{\ref{inst10}}$ \and A. A. C. Sander$^{\ref{inst3}}$ \and C. Erba$^{\ref{inst8}}$  \and V.~M.~A.~G\'omez-Gonz\'alez$^{\ref{inst1}}$ \and  C. Kehrig$^{\ref{inst7}}$ \and J. Klencki$^{\ref{inst12}}$ \and R. Kuiper$^{\ref{inst6}}$ \and A. Mehner$^{\ref{inst5}}$ \and S. E. de Mink$^{\ref{inst13}}$ \and M. S. Oey$^{\ref{inst4}}$  \and  V.~Ramachandran$^{\ref{inst3}}$ \and A. Schootemeijer$^{\ref{inst11}}$ \and S.~Reyero~Serantes$^{\ref{inst1}}$ \and A. Wofford$^{\ref{inst9}}$
 	}
 	
 	\authorrunning{D. Pauli et al.}
 	
 	\institute{
        Institut f{\"u}r Physik und Astronomie, Universit{\"a}t Potsdam, Karl-Liebknecht-Str. 24/25, 14476 Potsdam, Germany\label{inst1}
 	    \and Institute of Astronomy, KU Leuven, Celestijnenlaan 200D, 3001 Leuven, Belgium\label{inst2}
        \and Anton Pannekoek Institute for Astronomy, Science Park 904, 1098 XH Amsterdam, The Netherlands\label{inst10} 
 		\and Astronomisches Rechen-Institut, Zentrum für Astronomie der Universität Heidelberg, Mönchhofstr. 12-14, 69120 Heidelberg, Germany\label{inst3}
 		\and Department of Physics \& Astronomy, East Tennessee State University, Johnson City, TN 37614, USA\label{inst8}
 		\and Instituto de Astrofísica de Andalucía - CSIC, Glorieta de la Astronomía s.n., 18008 Granada, Spain\label{inst7}
        \and European Southern Observatory, Karl-Schwarzschild-Strasse 2, 85748 Garching bei München, Germany\label{inst12}
 		\and Faculty of Physics, University of Duisburg-Essen, Lotharstraße 1, D-47057 Duisburg, Germany\label{inst6}
        \and  Max Planck Institute for Astrophysics, Karl-Schwarzschild-Straße 1, D-85748 Garching, Germany\label{inst13}
 		\and ESO -- European Organisation for Astronomical Research in the Southern Hemisphere, Alonso de Cordova 3107, Vitacura, Santiago de Chile, Chile\label{inst5}
 		\and University of Michigan, 1085 South University Ave., Ann Arbor, MI 48109, USA\label{inst4}
        \and Argelander-Institut für Astronomie, Universität Bonn, Auf dem Hügel 71, 53121 Bonn, Germany\label{inst11}
        \and Instituto de Astronom\'ia, Universidad Nacional Aut\'onoma de M\'exico, Unidad Acad\'emica en Ensenada, Km 103 Carr. Tijuana$-$Ensenada, Ensenada, B.C., C.P. 22860, M\'exico\label{inst9}
 	}
 	
 	\date{Received ; Accepted}
 	
 	\abstract{
        The origin of the observed population of Wolf-Rayet (WR) stars in low-metallicity galaxies, such as the Small Magellanic Cloud (SMC), is not yet understood. Standard, single-star evolutionary models predict that WR stars should stem from very massive O-type star progenitors, but these are very rare. On the other hand, binary evolutionary models predict that WR stars could originate from primary stars in close binaries. 
        }
 	{
        We conduct an analysis of the massive O star, AzV\,14, to spectroscopically determine its fundamental and stellar wind parameters, which are then used to investigate evolutionary paths from the O-type to the WR stage with stellar evolutionary models. 
        }
        {
        Multi-epoch UV and optical spectra of AzV\,14 are analyzed using the non-local thermodynamic equilibrium (LTE) stellar atmosphere code PoWR. An optical TESS light curve was extracted and analyzed using the PHOEBE code. The obtained parameters are put into an evolutionary context, using the MESA code.
        }
        {
        AzV\,14 is a close binary system with a period of $P=\SI{3.7058\pm0.0013}{d}$. The binary consists of two similar main sequence stars with masses of $M_{1,2}\approx 32\,\msun$. Both stars have weak stellar winds with mass-loss rates of $\log{\dot{M}/(\msunpyr)}=-7.7\pm0.2$.
        Binary evolutionary models can explain the empirically derived stellar and orbital parameters, including the position of the AzV 14 components on the Hertzsprung-Russell diagram, revealing its current age of $\SI{3.3}{Myr}$. The model predicts that the primary will evolve into a WR star with $T_\mathrm{eff}\approx\SI{100}{kK}$, while the secondary, which will accrete significant amounts of mass during the first mass transfer phase, will become a cooler WR star with $T_\mathrm{eff}\approx\SI{50}{kK}$. Furthermore, WR stars that descend from binary components that have accreted significant amount of mass are predicted to have increased oxygen abundances compared to other WR stars. This model prediction is supported by a spectroscopic analysis of a WR star in the SMC.
 	}
        {
        Inspired by the binary evolutionary models, we hypothesize that the populations of WR stars in low-metallicity galaxies may have bimodal temperature distributions. Hotter WR stars might originate from primary stars, while cooler WR stars are the evolutionary descendants of the secondary stars if they accreted a significant amount of mass. These results may have wide-ranging implications for our understanding of massive star feedback and binary evolution channels at low metallicity. 
 	}
 	
 	\maketitle

 	\section{Introduction} 
 	\label{sec:intro} 
 	
        Massive stars ($M_{\rm initial}>8\,\msun$) in all evolutionary phases strongly affect their galactic neighborhood via stellar winds and ionizing fluxes. During core-H burning on the  main sequence, they have spectral types O and early B.  As massive stars evolve, their outer hydrogen (H) rich envelopes could be removed by stellar winds. The majority of massive stars are born in binary systems \citep[e.g][]{san2:12,san1:14,moe1:17}, and interactions with a companion star may also lead to a removal of the outer envelope. Highly evolved massive stars that lost a large portion of their outer H-rich layers and have optically thick winds are spectroscopically classified as Wolf-Rayet (WR) stars. The WR stars are typically hotter and have stronger, optically thick winds when compared with their evolutionary predecessors. These stars come in two major subtypes, WN and WC, which are spectroscopically identified by strong emission lines of nitrogen and carbon, respectively. WR stars end their lives with a core-collapse event, likely leading to the formation of black holes (BHs; e.g., \citealt{suk1:16,gal1:22}). Understanding the formation pathways of WR stars in nearby low-metallicity galaxies is needed  for quantifying stellar feedback and compact object populations in conditions resembling the early Universe.  
        
        The Small Magellanic Cloud (SMC) galaxy has a metallicity of ${Z\approx 1/7\,\zsun}$ \citep{hun1:07,tru1:07} and is nearby ($d\approx \SI{61}{kpc}$; \citealt{hil1:05}), thus providing an excellent test bed for investigating stars at low metallicity. The proximity of the SMC and its low foreground and intrinsic extinction allows for a detailed study of stellar and wind parameters of low-metallicity populations of O and WR stars. This enables us to obtain a realistic understanding of massive star evolution and feedback at low metallicity in general. 
 	     
        Yet what we have learned so far is perplexing. Only 12 WR stars exist in the SMC. All of them are very luminous and (except for one WO-type star) contain some hydrogen \citep{hai1:15,she1:16}. According to the standard, single-star evolutionary tracks, their progenitors are O-type  stars with masses  $\gtrsim40\,\msun$  \citep[][and references therein]{she1:20}. For an order of magnitude estimate, one can assume that a massive star spends $~10\,\%$ of its life in a WR stage. This implies that a galaxy containing approximately ten WR stars should contain approximately one-hundred massive O stars. However, there is a severe paucity of such O-stars in the SMC \citep{ram1:19,sch1:21}.

        Close binary evolution is expected to play a principal role in the formation of WR stars \citep{pac1:67,kip1:67}. However, the importance of binarity is still under debate and different studies have come to different conclusions \citep[e.g.,][]{Vanbeveren2007,eld1:17,she1:20,pau1:22}. Up to now, the evolution toward the WR stage in  binaries has chiefly been studied for primary stars, while the evolution of secondary stars has been largely neglected. In this paper, we aim to gain insights into the evolution of both components in a binary with well-established stellar parameters.

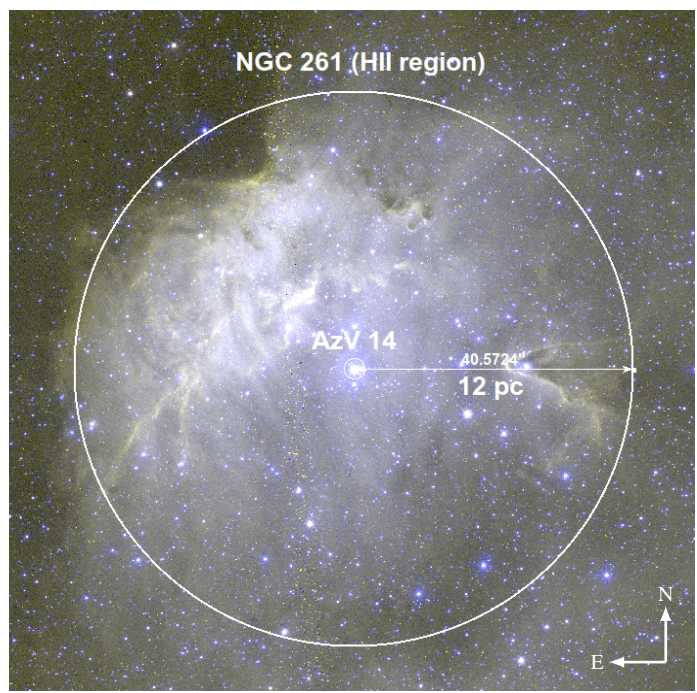
\begin{figure}[tpb]
     	    \centering
            \begin{tikzpicture}
                \node [anchor=south east] at (0,0)
         	    {\includegraphics[trim= 22.cm 5cm 21.cm 3cm ,clip ,width=0.49\textwidth]{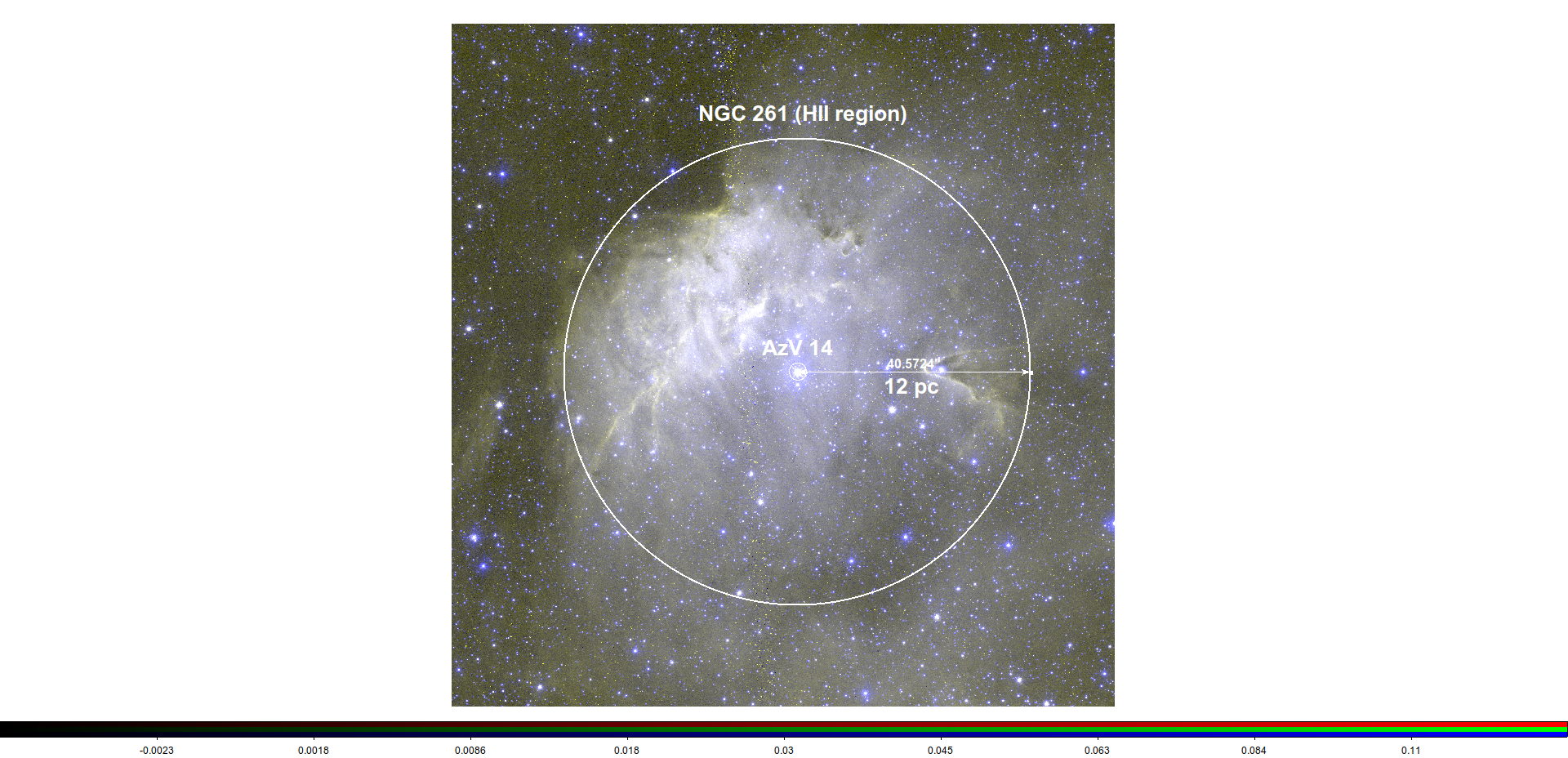}};
                \draw[white,line width=0.35mm, -latex] (-0.5,0.5) -- (-0.5, 1.25);
                \draw[white] (-0.5, 1.4) node {\tiny N};
                \draw[white,line width=0.35mm, -latex] (-0.5,0.5) -- (-1.25, 0.5);
                \draw[white] (-1.4, 0.5) node {\tiny E};
            \end{tikzpicture}
     	    \caption{False-color HST image composed of the images in the F475W (blue) and F657N (yellow) filters. The position of AzV\,14 and the approximate size of the NGC 261 \HII{} region are indicated by white circles.}
     	    \label{fig:HST_image}
     	\end{figure}

        One of the youngest and earliest O-type stars in the SMC is  AzV\,14, the main ionizing source of the \HII{} region NGC\,261 (see Fig.~\ref{fig:HST_image}). Previously, single epoch optical spectra of AzV\,14 have been analyzed by \citet{mas1:04} and \citet{mok1:06}. The star was classified as O5\,V  with a reported spectroscopic mass of $74\,\msun-90\,\msun$ being in the mass range of potential WR single-star progenitors. From a newly obtained light curve, we know now that AzV\,14 is in fact a binary. In this paper, we complement recently obtained
        spectra of AzV\,14 in the UV and optical by archival data, aiming at securely determining stellar parameters and, on this basis, to model the evolution of the AzV\,14 components toward WR stages.

 	\section{Observations}
 	\label{sec:Observations}
 	    
        Presently, six spectra of AzV\,14 obtained at different epochs covering the far-UV, UV and optical wavelength ranges exist in telescope archives. One far-UV spectrum was taken with the Far Ultraviolet Spectroscopic Explorer (FUSE; \citealt{oeg1:00}), two UV spectra with the Hubble Space Telescope's (HST) Faint Object Spectrograph (FOS; \citealt{key1:95}) and Space Telescope Imaging Spectrograph (STIS; \citealt{bra1:21}), and three optical spectra with the European Southern Observatory (ESO) Very Large Telescopes (VLT) Ultraviolet and Visual Echelle Spectrograph (UVES; \citealt{dek1:00}) and the X-Shooter \citep{ver1:11} spectrograph. One X-Shooter spectrum was taken as part of the XShootU project\footnote{\url{https://massivestars.org/xshootu/}} (from here on ``X-Shooter~(2020)'') and one as part of our program, ID 109.22V0.001 (from here on ``X-Shooter~(2022)''). A detailed description of the individual spectra and photometry can be found in Appendix~\ref{app:spectra}.

        AzV\,14 is close to the SMC bar, thus we adopt a distance modulus of ${DM=\SI{18.9}{mag}}$ \citep{wes1:97,har1:03,hil1:05}. 
        The radial velocities (RV) of the different regions in the SMC are not uniform \citep[e.g.,][]{pro1:10}. We estimate the RV shift of the NGC\,261 complex  by fitting Gaussians to interstellar absorption lines as ${\varv_\mathrm{NGC\,261}=\SI{148\pm2}{km\,s^{-1}}}$. All spectra shown in this work are in the rest frame of NGC\,261. 	    
        
        AzV\,14 was observed by the NASA Transiting Exoplanet Survey Satellite (TESS; \citealt{Ricker2015}) mission (TIC~180206579) during its sectors 1, 27 and 28 in full-frame image (FFI) mode with a cadence of $30$, $10$ and $\SI{10}{min}$, respectively. TESS has a low spatial resolution of ${21\si{\arcsec\,px^{-1}}}$. However, since AzV\,14 is the brightest optical object in this region we can extract its light curve (see Appendix~\ref{sec:tess}). Furthermore,  AzV\,14 was detected by the \textit{Chandra} X-ray telescope \citep{cxo2000}.

 	    \begin{figure*}[tbp]
 	        \centering
 	        \includegraphics[width=\textwidth]{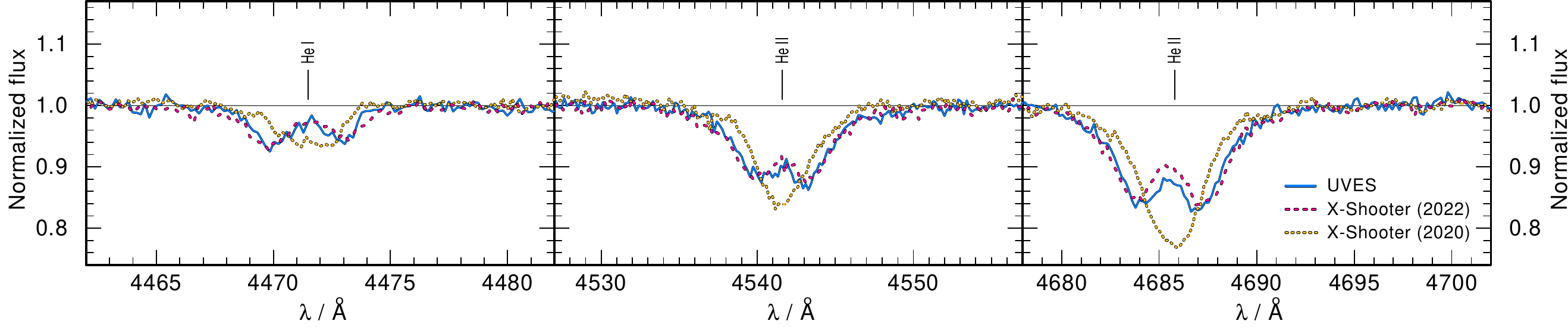}
 	        \caption{Comparison of selected lines in the UVES (solid blue) and the X-Shooter (dashed pink and dotted yellow) optical spectra. The high resolution UVES spectrum is binned by $\SI{0.15}{\AA}$ to match the X-Shooter spectra. All spectra are convolved with a Gaussian having a $\mathrm{FWHM}=\SI{0.05}{\AA}$ for a better comparison. The lines in the UVES and X-Shooter~(2022) spectra are separated, while they are blended in the X-Shooter~(2020) spectrum. This provides a clear indication for binarity.}
 	        \label{fig:comparison_spectra}
 	    \end{figure*}

    \section{Spectral analysis}
    \label{sec:spectral_analyis}
    \subsection{Spectral line variability reveals AzV\,14 as a close binary  }
    \label{sec:compare_spectra}
 	
        The multi-epoch spectroscopy allows us to search for signs of binarity. In Fig.~\ref{fig:comparison_spectra}, we show selected \HeI{} and \HeII{} lines in the UVES and X-Shooter spectra, taken at different epochs. \citet{mas1:04} and \citet{mok1:06} attributed the apparent emission feature in the middle of helium absorption lines of the UVES to a nebular contamination. New multi-epoch X-Shooter spectra clearly show that the profiles of helium lines are complex and variable. They are well explained as composite lines originating from two components of a binary system. The UVES and X-Shooter~(2022) spectra must have been taken at a comparable orbital phase, as the line profile is similar in both spectra, while the X-Shooter~(2020) spectrum must have been obtained close to conjunction. This is confirmed by the binary ephemeris derived in the Sect.~\ref{sec:phoebe}. 

    \subsection{Establishing projected rotation velocity and RVs}
    
        The projected rotation velocity as well as the RV shifts of the individual binary components are necessary ingredients of spectral modeling. In the following we describe how we derived these quantities.
        
    \subsubsection{Projected rotation velocities}

        The helium lines in UVES and X-Shooter (2022) spectra are resolved into two components with lines of similar depths and widths, indicating that the components of the binary system are similar stars. We employ the \verb+iacob broad+ tool \citep{sim1:14} to measure the projected rotational velocities for each binary component. For the fitting procedure the lines \HeII{}~$\lambda4200$, \HeI{}~$\lambda4471$, \HeII{}~$\lambda4542$, and \HeII{}~$\lambda4686$ were used, while cutting off the parts of the lines that are blended by the companion. Although helium lines are not only rotationally, but also pressure broadened, they were analyzed because of the lack of available metal lines. The \verb+iacob broad+ tool derives two different values for the projected rotational velocities, one from a Fourier transformation (FT) and one from the goodness of the fit (GoF). We quote the average value obtained by the different methods. As the estimates on the projected rotation velocities are based on blended lines, we adopt conservative error margins. We find that both binary components have similar projected rotational velocities of ${\varv_\mathrm{rot}\,\sin i=\SI{90\pm20}{km\,s^{-1}}}$.

    \subsubsection{RVs of the binary components in each spectrum}
    
        The RVs of all spectra are determined by shifting the synthetic spectra until they fit the observations using a Markov chain Monte Carlo method combined with a least square fitting \citep[][their section 3.1.2 and their appendix A]{pau2:22}.

        The UVES spectrum shows RVs of ${\varv_1=\SI{-101.9\pm2.6}{km\,s^{-1}}}$ and ${\varv_2=\SI{80.5\pm3.2}{km\,s^{-1}}}$ for the primary and the secondary, respectively. This is comparable to those of the X-Shooter~(2022) spectrum, yielding RV shifts of ${\varv_1=\SI{-112.4\pm2.6}{km\,s^{-1}}}$ and ${\varv_2=\SI{98.6\pm2.9}{km\,s^{-1}}}$. In the X-Shooter~(2020) spectrum the lines are not split due to an orbital phase close to conjunction. We estimated ${\varv_1=\SI{-46.1\pm5.8}{km\,s^{-1}}}$ and ${\varv_2=\SI{22.3\pm5.9}{km\,s^{-1}}}$ for the primary and secondary, respectively. Accurate measurements of RV shifts in the optical spectra allow us to estimate the temperature and surface gravity of the individual binary components (see Sect.~\ref{sec:t_and_logg}).
        
        In the STIS spectrum, the \HeII{}\,$\lambda1640$ line is split into two individual components of similar strength (see Fig.~\ref{fig:wind}), yielding RVs of  ${\varv_1=\SI{-128.6\pm2.5}{km\,s^{-1}}}$ for the primary and ${\varv_2=\SI{113.2\pm2.3}{km\,s^{-1}}}$ for the secondary. Measuring RVs in the UV spectra allows to determine the terminal wind velocity ($\varv_\infty$) and thus the mass-loss rate ($\dot{M}$) more precisely  (see Sect.~\ref{sec:wind}). We estimated the RVs in the low resolution FOS spectrum such that the width of the oxygen lines in the range of $\SIrange{1330}{1420}{\AA}$ can be matched, arriving at  ${\varv_1=\SI{92.9\pm6.2}{km\,s^{-1}}}$ for the primary and ${\varv_2=\SI{113.2\pm2.8}{km\,s^{-1}}}$ for the secondary.

    \subsection{Stellar atmosphere modeling}
    \label{sec:PoWR}        
    \label{sec:spectral_fitting}
 	
        To analyze the spectra, we employ the Potsdam Wolf-Rayet (PoWR) model atmosphere code \citep{gra1:02,ham1:03,ham1:04,tod2:15,san2:15}. A short characterization of the code is given in Appendix~\ref{app:PoWR}.
        
    \subsubsection{Temperatures and surface gravities}
    \label{sec:t_and_logg}
    
        We measure the stellar temperatures using the ratio between \HeI{} to \HeII{} lines (see Fig.~\ref{fig:H_and_He_lines}). To constrain the surface gravities, we used the wings of the Balmer, \Hbeta{}, \Hgamma{}, and \Hdelta{} lines in the UVES and X-Shooter~(2022) spectra. For the primary we obtained an effective temperature of ${T_1=\SI{43\pm2}{kK}}$, while the secondary is slightly cooler with ${T_2=\SI{42\pm2}{kK}}$. The surface gravity is ${\log g = 4.0\pm0.2}$ for both binary components. 
 	
    \subsubsection{Stellar luminosities and spectroscopic masses}
 	
        Luminosity $L$ and color excess $E_\mathrm{B-V}$ are determined by fitting the composite spectral energy distribution (SED) --- containing the synthetic flux of both stellar components --- to photometry (see Fig.~\ref{fig:SED}). The color excess is modeled as a combined effect of Galactic foreground, using the reddening law by \citet{sea1:79} with ${E_\mathrm{B-V,\,Gal}=\SI{0.03}{mag}}$, and SMC background, using the reddening law by \citet{bou1:85} with ${E_\mathrm{B-V,\,SMC}=\SI{0.11}{mag}}$.
        
        The luminosities of the two binary components are chosen such that the ratio of the depths of the synthetic helium lines match all spectra. Additionally, we use the \CIII{}\,$\lambda1175$ line in the UV, which is sensitive to changes in the light ratio. Both stars have comparable luminosities in the optical and UV. The final luminosities are ${\log (L_1/\lsun) = 5.41\pm0.15}$ and ${\log (L_2/\lsun) = 5.38\pm0.15}$ for the primary and secondary, respectively. Accordingly, the masses of the primary and secondary are  ${M_\mathrm{spec,\,1}=32^{+8}_{-7}\,\msun}$ and ${M_\mathrm{spec,\,2}=31_{-7}^{+8}\,\msun}$. The spectroscopic mass ratio is $q_\mathrm{spec}=0.97$. When employing the method of \citet{wil1:41}, one can derive an independent measure of the mass ratio from the RVs. This yields $q_\mathrm{Wilson}=0.95\pm0.02$ which is in agreement with the spectroscopic results.

    \subsubsection{CNO surface abundances}
 	
        All metal lines are well-matched with standard initial abundances of the SMC  (see Table~\ref{tab:elements} and Fig.~\ref{fig:metal_lines}). The only line which is not well reproduced is the \NIV{}\,$\lambda3479$ line, which is deeper than predicted by the synthetic spectrum. This absorption line is quite narrow, suggesting that the \NIV{}\,$\lambda3479$ line might be blended with an unidentified ISM line. 
        Pristine abundances prompt us to conclude that the AzV\,14 components are young unevolved stars which have not yet interacted.

        \begin{table}[tbp]
            \centering
            \caption{Summary of the stellar parameters of both binary components obtained from spectroscopic analysis conducted with the PoWR code.}
            \begin{tabular}{lccc}\hline \hline \rule{0cm}{2.8ex}%
                \rule{0cm}{2.2ex} parameter & primary                 & secondary  & unit \\
                \hline \rule{0cm}{3.4ex}%
                \rule{0cm}{2.8ex}$T_\mathrm{eff}^{\,\,\,\,(a)}$         &   $42.8\pm2.0$       &   $41.8\pm2.0$           & $[\si{kK}]$ \\
                \rule{0cm}{2.8ex}$\log\,g$           &$\,\,4.0\pm0.2$ &$\,\,4.0\pm0.2$ & $[\si{cm\,s^{-2}}]$ \\
                \rule{0cm}{2.8ex}$\log\,L$                     &  $5.41\pm0.15$  &   $5.38\pm0.15$ & $[\lsun]$\\
                \rule{0cm}{2.8ex}$R$                           & $9.3\pm0.5\,$ & $\,\,9.2\pm0.5$ & $[\rsun]$\\
                \rule{0cm}{2.8ex}$R_\mathrm{RL}^{\,\,\,\,(b)}$                           & $15.6\pm0.8\,$ & $\,\,15.2\pm0.7$ & $[\rsun]$\\
                \rule{0cm}{2.8ex}$R/R_\mathrm{RL}$                           & $0.60\pm0.04$ & $0.61\pm0.04$ & \\
                \rule{0cm}{2.8ex}$M_\mathrm{spec}$                           &   $\,\,32^{+8}_{-7}$ &  $\,\,31^{+8}_{-7}$  & $[\msun]$ \\
                \rule{0cm}{2.8ex}$\log \dot{M}$             &  $-7.7\pm0.2\,$&   $-7.7\pm0.2\,$ & $[\msunpyr]$\\
                \rule{0cm}{2.8ex}$\varv_\infty$      &$1600\pm200$    &$1600\pm200$        & $[\si{km\,s^{-1}}]$\\
                \rule{0cm}{2.8ex}$\varv_\mathrm{rot} \sin i^{\,\,\,\,(c)}$      &   $90\pm20$    &   $90\pm20$ & $[\si{km\,s^{-1}}]$ \\
                \rule{0cm}{2.8ex}$\varv_\mathrm{rot} ^{\,\,\,\,(d)}$      &   $157\pm40$    &   $157\pm40$  & $[\si{km\,s^{-1}}]$\\
                \rule{0cm}{2.8ex}$P_\mathrm{rot} ^{\,\,\,\,(d)}$      &   $3.0\pm0.8$    &   $3.0\pm0.8$  & $[\si{d}]$\\
                \rule{0cm}{2.8ex}$\log\,Q_\ion{H}\,$     &   $49.13$              &   $49.10$   &       $[\si{s^{-1}}]$\\                
                \rule{0cm}{2.8ex}$\log\,Q_{\HeI{}}\,$    &   $48.41$              &   $48.35$     &  $[\si{s^{-1}}]$   \\                
                \rule{0cm}{2.8ex}$\log\,Q_{\HeII{}}\,$   &   $43.41$              &   $43.27$   &    $[\si{s^{-1}}]$\vspace{1ex} \\
                \hline
            \end{tabular}
            \rule{0cm}{2.8ex}%
            \begin{minipage}{0.95\linewidth}
                \ignorespaces 
                 $^{(a)}$ For a definition, see Appendix~\ref{app:PoWR}. $^{(b)}$ Calculated using the orbital parameters obtained with the PHOEBE code (see Sect.\,\ref{sec:phoebe}) and the formula of \citet{egg:83}. $^{(c)}$ Obtained with the \verb|iacob broad| tool. $^{(d)}$ Based on the inclination $i=35^\circ$ obtained with the PHOEBE code (see Sect.\,\ref{sec:phoebe}).
            \end{minipage}
            \label{tab:stellar_parameters_summary}
        \end{table}
        
 	\subsubsection{Wind mass-loss rates}
        \label{sec:wind}
 	
        In the optical spectra of AzV\,14, \Halpha{} and \HeII{}\,$\lambda4686$ do not show any contribution from winds. However, in the UV one can see wind lines with P\,Cygni profiles. Key diagnostic lines are the \NV{}\,$\lambda\lambda1239, 1243$ and \CIV{}\,$\lambda\lambda1548, 1551$ resonance doublets and the \HeII{}\,$\lambda1640$ line. By fitting the observed \NV{} and \CIV{} resonance doublets in the STIS and FOS spectra consistently yields a terminal wind velocity of ${\varv_\infty=\SI{1600\pm200}{km\,s^{-1}}}$ and a mass loss rate of ${\log(\dot{M}/(\msunpyr))=-7.7}$ for each of the stars (see Fig.~\ref{fig:wind}). 
        
        To match the \OVI{}\,$\lambda\lambda1032,1038$ resonance doublet  in the FUSE spectrum, we include in the model a hot plasma component with ${T_\mathrm{X}=\SI{3}{MK}}$ emitting X-rays with ${L_\mathrm{X,mod}=\SI{2e30}{erg\,s^{-1}}}$. 
        According to the {\em Chandra} Point Source Catalog v.2.0 \citep{CSC2010}, the X-ray flux of AzV\,14  in the $\SIrange{0.5}{7.0}{keV}$ band is ${F_\mathrm{X}=\SI{7.3e-16}{erg\,s^{-1}\,cm^{-2}}}$, corresponding to an X-ray luminosity corrected for reddening (see Fig.\,\ref{fig:SED})  ${L_\mathrm{X}\approx \SI{5e32}{erg\,s^{-1}}}$ or ${\log(L_\mathrm{X}/L_\mathrm{bol})\approx -6}$.
        We suggest that the majority of the observed X-rays originate from a colliding wind zone between two binary components in common with O-type binaries in the Galaxy \citep{rau1:16}. A summary of the empirically determined stellar parameters is given in Table~\ref{tab:stellar_parameters_summary}.

     	\begin{figure}[tpb]
     	    \centering
     	    \includegraphics[width=0.48\textwidth]{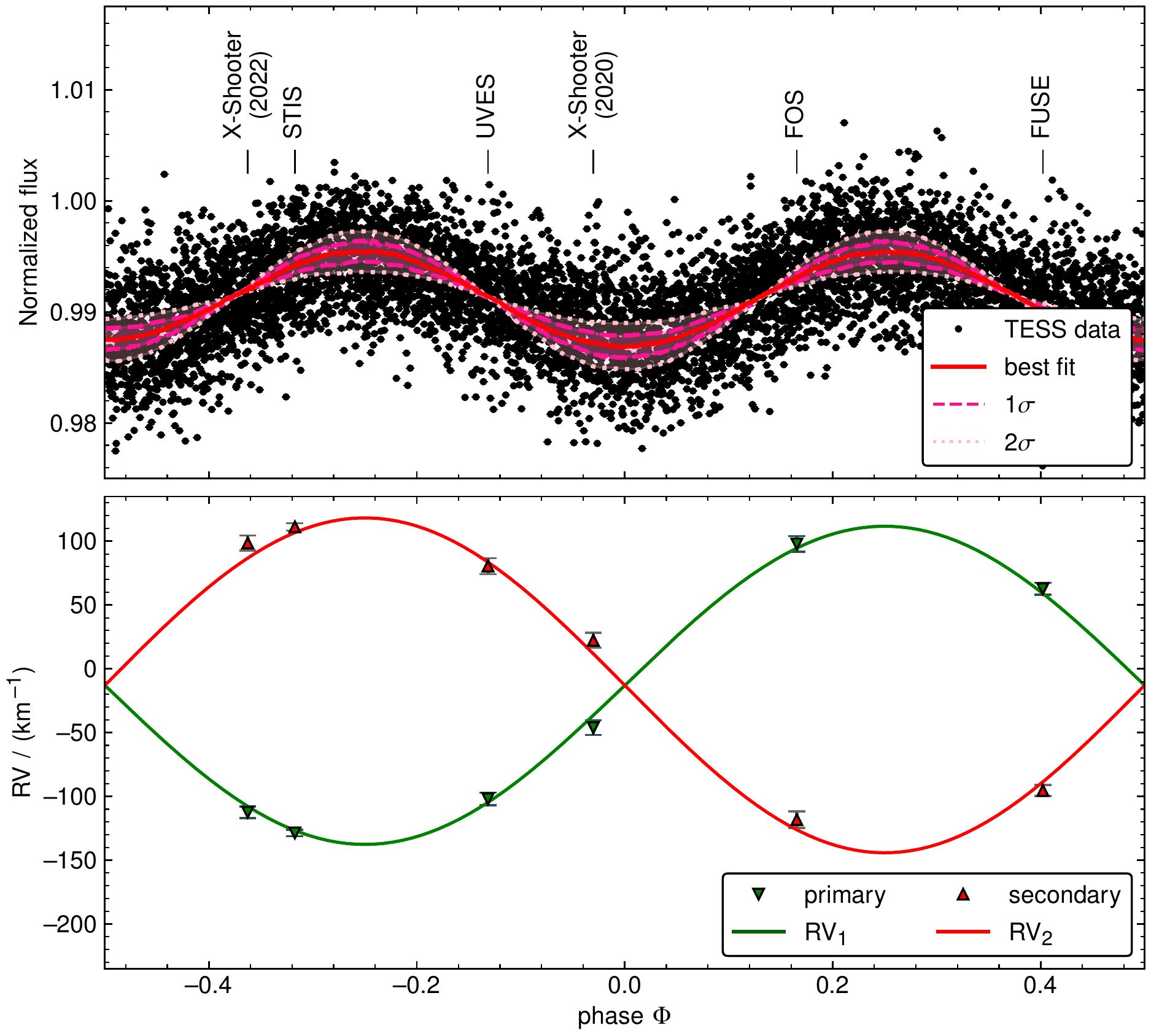}
     	    \caption{Comparison of the phased observed and synthetic light and RV curve of AzV\,14. \textit{Upper panel}: Phased TESS light curve (dots) and best fit obtained with the PHOEBE code (red line). For the PHOEBE model we show the $1\sigma$ and $2\sigma$ deviations as pink shaded areas. \textit{Lower panel}: Observed (triangles) and fitted RV curves (solid lines) of the primary (green) and secondary (red) component.}
     	    \label{fig:phoebe}
     	\end{figure}

    \section{Orbital analysis}
    \label{sec:phoebe}

     	\begin{figure*}[thpb]
     	    \centering
     	    \includegraphics[trim= 1.5cm 1.cm 1.5cm 0cm ,clip ,width=\textwidth]{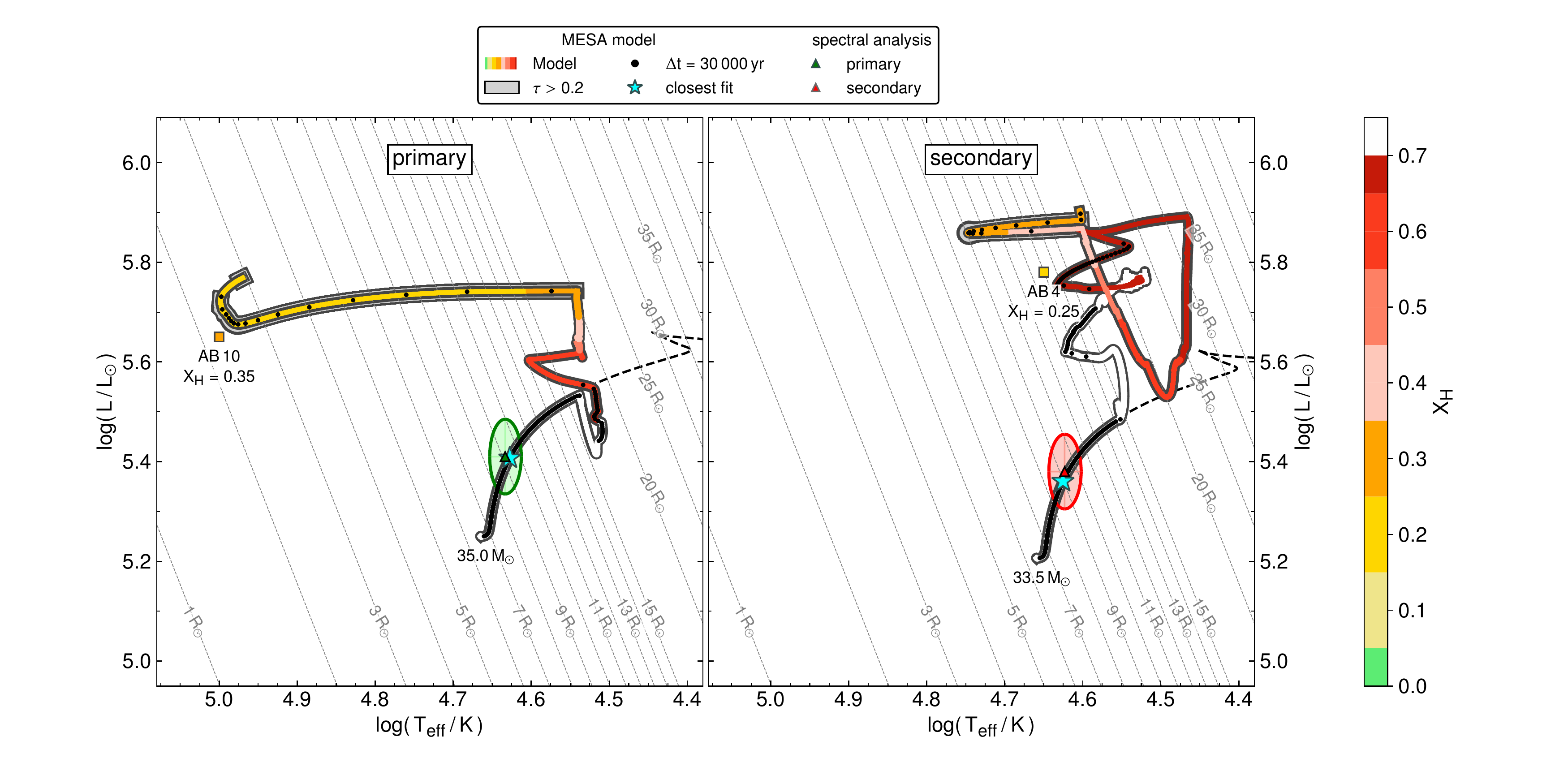}
     	    \caption{Evolutionary tracks of the primary (left) and secondary (right) compared to the empirical positions of AzV\,14 (triangles with error ellipses). The evolutionary tracks are color-coded by the surface hydrogen abundance of the model and are overlayed by black dots corresponding to equidistant time-steps of $\SI{0.3}{Myr}$. In the background, single star tracks are shown as dashed black lines. The instance of time in which the model is closest to the observations is marked by blue stars. The tracks are labeled by their initial stellar masses. The phases during which the model is expected to be observed as WR star, this is where $\tau\geq0.2$, are highlighted by bold black frames. In addition, the positions of the WR stars AB\,10 and AB\,4, are marked by squares which are color-coded according to their observed surface-hydrogen abundances.}
     	    \label{fig:mesa_best_fit}
     	\end{figure*}

        The TESS light curve of AzV\,14 displays periodic variability (originating from ellipsoidal variability, see Fig.~\ref{fig:phoebe}). We employed a frequency analysis and phase folding techniques to determine the dominant periodicity in the TESS light curve, yielding an orbital period of ${P_\mathrm{orb}=\SI{3.7058\pm0.0013}{d}}$ (see Appendix~\ref{sec:tess}). 

        Orbital parameters from the light and RV curve are determined consistently with the Physics of Eclipsing Binaries (PHOEBE) code version 2.4.5 \citep{prs1:05,prs1:11,prs1:16,hor1:18,jon1:20,con1:20}. The input parameters are provided by the spectral analysis (Table~\ref{tab:stellar_parameters_summary}). To reduce the free parameter space, the orbital period is fixed to the period measured from the TESS light curve. Since the light curve is sinusoidal and has minima at phases $\Phi=0.0$ and $\Phi=0.5$, the binary orbit must be close to circular ($e=0$). We fit the remaining orbital parameters, including the inclination. A more detailed description can be found in Appendix~\ref{app:phoebe}. 
        
        The best fit is achieved with an inclination of ${i=\SI{35\pm5}{^\circ}}$ and an epoch of the primary eclipse of ${T_{0}=\SI{2459036.101 \pm 0.004}{HJD}}$. The orbital masses are $M_\mathrm{orb,\,1}=33.6^{+5.0}_{-3.7}$ and $M_\mathrm{orb,\,1}=31.9^{+4.8}_{-3.5}$, being in agreement with the spectroscopic analysis. The remaining orbital parameters are listed in Table~\ref{tab:phoebe_results1} and \ref{tab:phoebe_results2}. The best fitting light and RV curves are shown in Fig.~\ref{fig:phoebe}, including the phases at which each spectrum was taken. The X-Shooter~(2020) spectrum was obtained in the orbital phase close to conjunction, while the UVES and X-Shooter~(2022) spectra were measured at comparable phases out of conjunction.

        Given the inclination of the system, the rotation velocities of the two components are $\varv_\mathrm{rot}=\SI{157\pm40}{km\,s^{-1}}$. Hence, the rotation period of each star is $P_\mathrm{rot}=\SI{3.0\pm0.8}{d}$ which is close to the orbital period, implying that the binary is tidally locked.

        Using the orbital period and the mass-ratio, we calculated the Roche radii for each star (see Table~\ref{tab:stellar_parameters_summary}). By comparison to the previously determined stellar radii we conclude that both stars are underfilling their Roche lobe ($R/R_\mathrm{RL}=0.6$), further supporting our conclusion that both stars have not interacted yet.

 	\section{Binary evolution models predict the formation of WR stars}
 	\label{sec:evol_analysis}

        Empirically derived stellar parameters  of the binary components  of AzV\,14 (Table~\ref{tab:stellar_parameters_summary}) are used to anchor the position of the system on the Hertzsprung–Russell diagram (HRD), and on this basis to investigate their possible future evolution. We calculated binary evolution models with the Modules for Experiments in Stellar Astrophysics (MESA) code \citep{pax1:11,pax1:13,pax1:15,pax1:18,pax1:19}. The methods and basic assumptions made in our models are described in Appendix~\ref{sec:input_physics}. In our models, a star enters WR evolutionary stages when the optical depth at its surface is $\tau\geq0.2$. For further information we refer to Appendix~\ref{app:WR_phase}.
 	    
        The current positions of the binary components in the HRD can be explained best by a binary evolutionary model with initial masses of ${M_\mathrm{1,ini}=35.0\,\msun}$ and ${M_\mathrm{2,ini}=33.5\,\msun}$, and an initial period of ${P_\mathrm{orb,ini}=\SI{3.7}{d}}$.  The model predicts that AzV\,14 formed $\SI{3.3}{Myr}$ ago. The corresponding evolutionary tracks are shown in Fig.~\ref{fig:mesa_best_fit}. According to the evolutionary model the binary components have not interacted yet,  but will exchange mass in the future. During the future mass-transfer event, it is predicted that the secondary will accrete about $~15\,\msun$. Both binary components evolve successively into WR stars.

     	\begin{table}[t]
     		\centering
     		\caption{Summary of the stellar parameters of both binary components reproduced with the MESA stellar evolution code.}
     		\begin{tabular}{lccc}\hline \hline \rule{0cm}{2.8ex}%
     			\rule{0cm}{2.2ex} parameter & primary                 & secondary  & unit \\
     			\hline \rule{0cm}{3.4ex}%
     			\rule{0cm}{2.8ex}$T_\mathrm{eff}$         &   $42.5$       &   $42.2$           & $[\si{kK}]$ \\
     			\rule{0cm}{2.8ex}$\log\,g$           &$4.04$ &$4.05$ & $[\si{cm\,s^{-2}}]$ \\
     			\rule{0cm}{2.8ex}$\log\,L$                     &  $5.41$  &   $5.36$ & $[\lsun]$\\
     			\rule{0cm}{2.8ex}$R$                           & $9.5$ & $9.1$ & $[\rsun]$\\
                \rule{0cm}{2.8ex}$M_\mathrm{ini}$                           &   $35.0$ &  $33.5$  & $[\msun]$ \\
     			\rule{0cm}{2.8ex}$M_\mathrm{evo}$                           &   $33.7$ &  $32.4$  & $[\msun]$ \\
     			\rule{0cm}{2.8ex}$\log \dot{M}^{(a)}$             &  $-6.5$&   $-6.6$ & $[\msunpyr]$\\
     			\rule{0cm}{2.8ex}$\varv_\mathrm{rot}$      &   $129$    &   $123$  & $[\si{km\,s^{-1}}]$\vspace{1ex} \\
     			\hline
     		\end{tabular}
     		\rule{0cm}{2.8ex}%
     		\begin{minipage}{0.95\linewidth}
     			\ignorespaces 
     			 $^{(a)}$ According to the mass-loss recipes used in our evolutionary models (see Appendix~\ref{sec:input_physics}).
     		\end{minipage}
     		\label{tab:stellar_parameters_summary_MESA}
     	\end{table}

        Both components will become H-poor WN type stars similar to WN stars observed in the SMC. For comparison the HRD shown in Fig.\,\ref{fig:mesa_best_fit} includes the positions of the apparently single WR stars SMC AB\,10 and AB\,4. The primary will enter the WN stage at an age of $\SI{5.9}{Myr}$ and spends most of its WR lifetime ($\SI{0.35}{Myr}$) close to the helium zero-age main sequence, which is roughly at ${\log(T_\mathrm{eff}/\si{K})\approx5.0}$  (SMC AB\,10 matches this position).  During the WR stage, the primary will have a mass of $17\,\msun$, while being accompanied by a $~50\,\msun$ main sequence star in an $\approx\SI{8.5}{d}$ orbit. The mass ratio has only a second-order effect on the amount of mass removed during the mass-transfer phase \citep[e.g.,][their section 3.2]{pau1:22}, meaning that the primary would be located (if it goes through a stable mass-transfer event) at a comparable position in the HRD, even if it had a less massive companion that could prevent or complicate a detection. We presume that at the end of its evolution, the primary will directly collapse into a BH and remains bound.

        The remaining secondary star, will continue its evolution on the main sequence.  Shortly after that stage, it will initiate mass-transfer onto the BH, stripping off parts of its 
        H-rich envelope and entering the WR phase at an age of $\SI{6.9}{Myr}$. The resulting secondary WR star will spend most of its lifetime (also $\SI{0.35}{Myr}$) at much lower temperatures (${\log(T_\mathrm{eff}/\si{K})\approx4.75}$) than the primary WR star while being similarly massive with $~25\,\msun$ (SMC AB\,4 is at this position). This relatively cool WR star is accompanied by a BH with a mass of $~16\,\msun$ in an orbit of $\SI{3.3}{d}$.

\begin{figure}[tpb]
            \centering
            \includegraphics[trim= 1.1cm 0cm 0.cm 0cm ,clip ,width=0.48\textwidth]{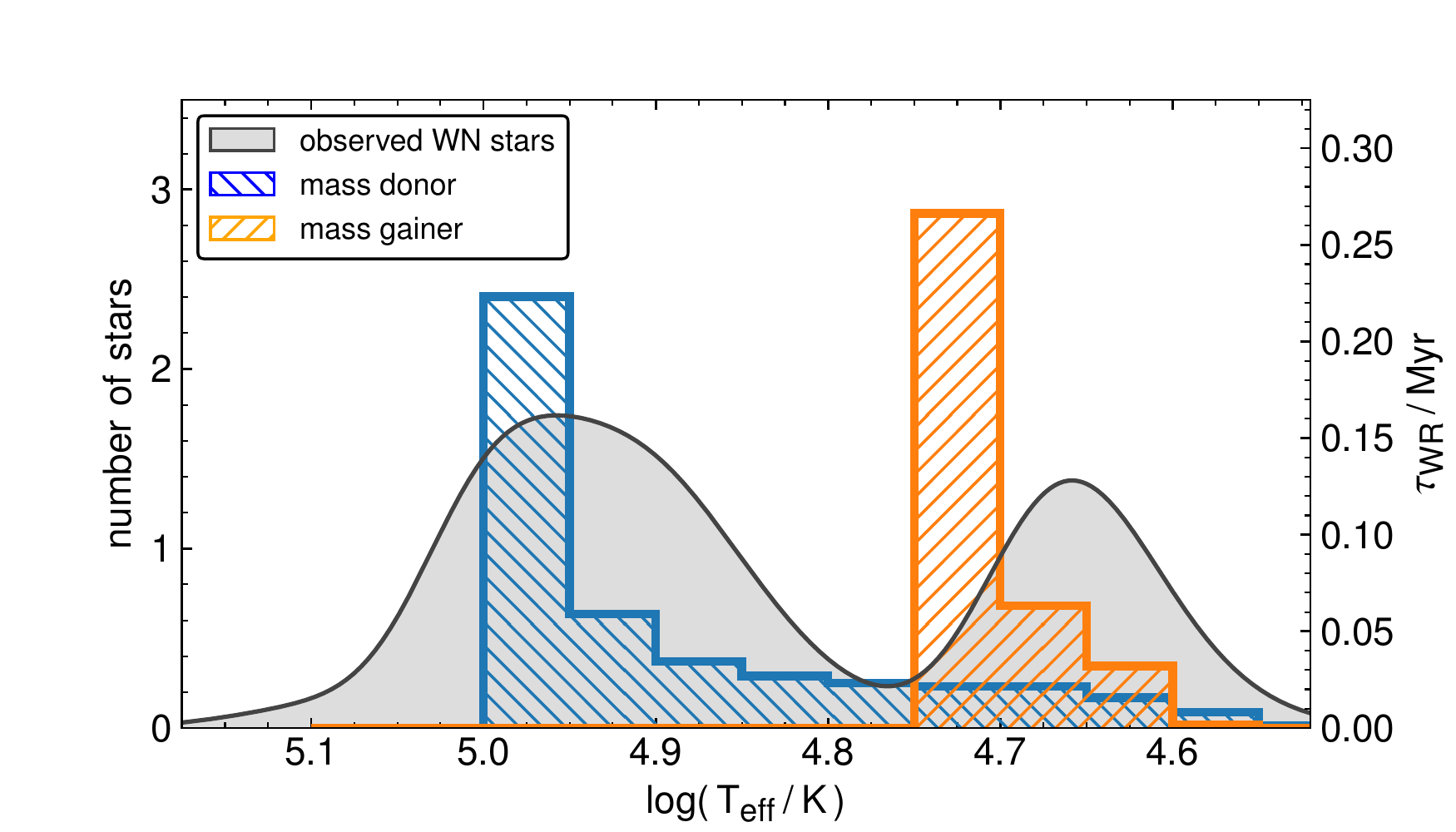}
            \caption{Observed temperature distribution of the WR stars in the SMC (gray area) compared to the time our evolutionary model of the primary (blue) and secondary (orange) are predicted to spend in the different temperature ranges during their WR phase. The observations are shown as Gaussians that have standard deviations corresponding to the observational uncertainties (see Table~\ref{tab:WR_stars}). We excluded the binary SMC AB5 from this plot, as it has a different evolutionary origin.}
            \label{fig:WR_dist}
        \end{figure}

        Our binary evolutionary models of AzV\,14 predict the formation of WR stars with different temperatures. Indeed, the WR population in the SMC \citep{hai1:15,she1:16,she1:18} has a bimodal temperature distribution which is comparable to the predicted temperature regimes of the evolutionary models of AzV\,14 (see Fig.~\ref{fig:WR_dist}). In the following, we elaborate on this idea further and discuss the implications and robustness of our findings.

    \section{Discussion}
    \label{sec:discuss}
    
    \subsection{Exploring the parameter space by computing a small grid of binary evolutionary models}
    \label{app:mesa_grid}
        
        \begin{figure*}[thpb]
            \centering
            \includegraphics[trim= 2.0cm 0cm 5.5cm 0cm ,clip ,width=\textwidth]{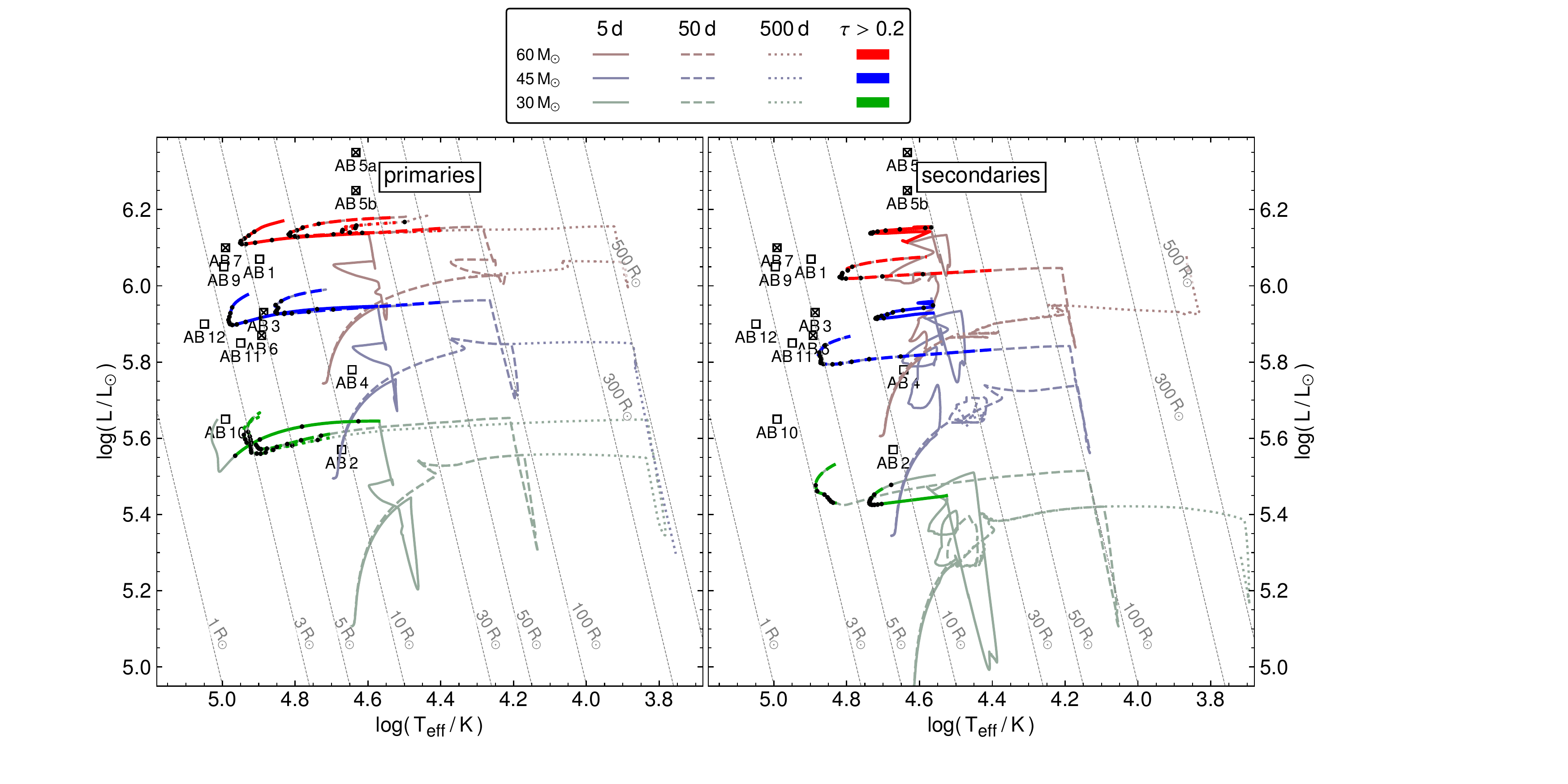}
            \caption{Evolutionary tracks of primary (left) and secondary (right) models with initial primary masses of $30$ (green gray), $45$ (blue gray), and $60\,\msun$ (red gray), fixed mass ratio $q=0.85$ and initial orbital periods of $5$ (solid), $50$ (dashed), and $\SI{500}{d}$ (dotted). We highlighted the WR phase (i.e., $\tau\geq0.2$) in bold colors. Equidistant time steps of $\SI{0.3}{Myr}$ are marked by black dots during the WR stage. In the background, we marked the positions of all observed WR stars in the SMC.}
            \label{fig:mesa_grid}
        \end{figure*}

        Inspired by the insights gained from the binary evolutionary models of AzV\,14 (see Sect.~\ref{sec:evol_analysis}), we  explore if the bimodal temperature distribution of WR stars at low metallicity is also predicted by binary evolutionary models with different initial masses and initial orbital periods. Therefore, we calculated a small grid covering initial primary masses of $30$, $45$, and $60\,\msun$, while having a fixed mass ratio of $q=0.85$. The initial primary masses are chosen to roughly represent the full luminosity distribution of the observed WR population of the SMC. The models are calculated for initial orbital periods of $5$, $50$, and $\SI{500}{d}$. A HRD containing the evolutionary tracks of all the models is shown in Fig.~\ref{fig:mesa_grid}.

    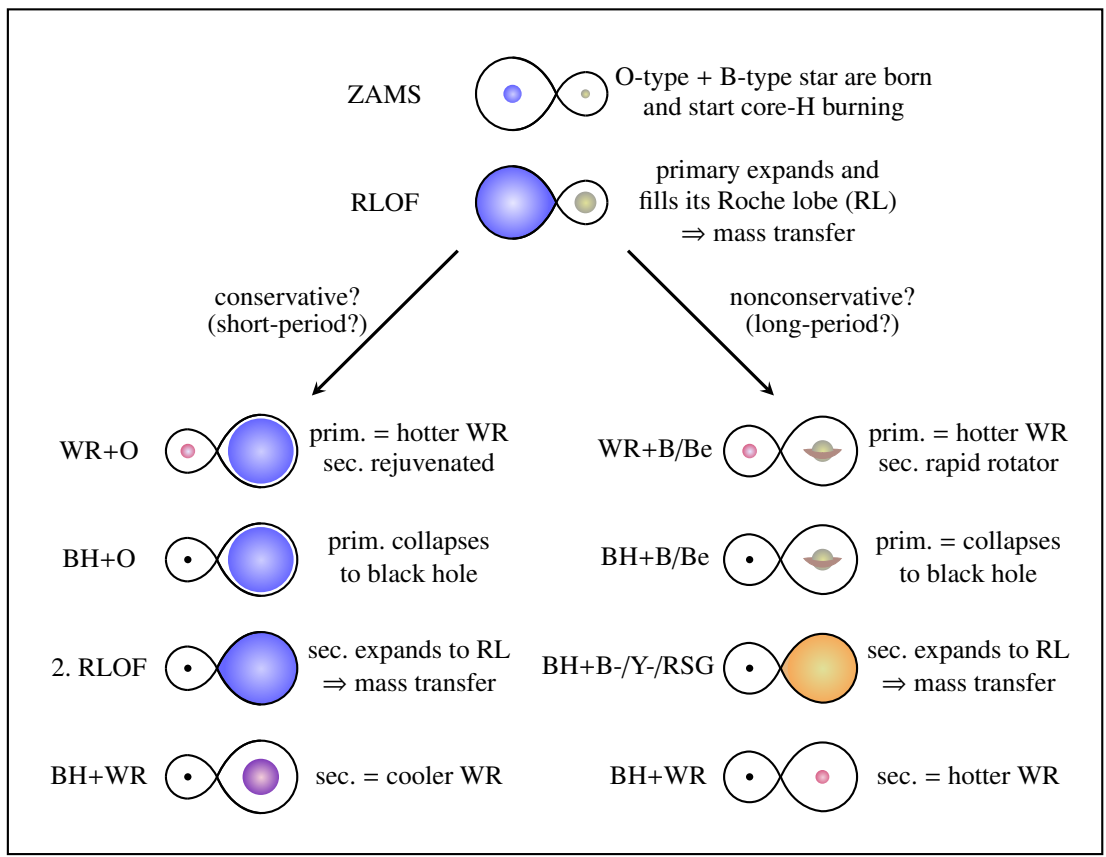
\begin{figure*}[pth]
        \centering
        \begin{minipage}[c]{0.79\textwidth}
        
            \begin{tikzpicture}[scale = 1.6]
                \draw[black,thick] (-4,-4.8) rectangle (5.,2.2);
                \begin{scope}[scale = 0.6, xshift = 0.25cm,yshift = 2.5cm]
                    \draw[thick] (0,0.5) arc (90:180:0.5cm) arc (-180:-90:0.5cm) .. controls (0.4,-0.5) and (0.6,0) .. (0.6,0) .. controls (0.4,0.5) and (0,0.5) .. (0,0.5);
                    \draw[thick] (1.3,0) arc (0:90:0.3cm);
                    \draw[thick] (1.3,0) arc (0:-90:0.3cm);
                    \draw[thick] (0,-0.5) .. controls (0.4,-0.5) and (0.6,0) .. (0.6,0) .. controls (0.75,0.3) and (1,0.3) .. (1.,0.3);
                    \draw[thick] (0,0.5) .. controls (0.4,0.5) and (0.6,0) .. (0.6,0) .. controls (0.75,-0.3) and (1,-0.3) .. (1.,-0.3);

                    \shade[outer color = blue!60!white, inner color = blue!20!white] (0,0) circle (0.1/0.8);
                    \shade[outer color = blue!60!white!60!yellow, inner color = blue!20!white!60!yellow] (1.,0) circle (0.05/0.8);
                \end{scope}

                \begin{scope}[scale = 0.6, xshift = 0.25cm,yshift = 1.cm]
                    \filldraw[color=blue!60!white] (0,0.5) arc (90:180:0.5cm) arc (-180:-90:0.5cm) .. controls (0.4,-0.5) and (0.6,0) .. (0.6,0) .. controls (0.4,0.5) and (0,0.5) .. (0,0.5);
                    \shade[outer color = blue!60!white, inner color = blue!10!white] (0,0) circle (0.5);
                    \shade[outer color = blue!60!white!60!yellow, inner color = blue!20!white!60!yellow] (1,0) circle (0.15);
                    
                    \draw[thick] (0,0.5) arc (90:180:0.5cm) arc (-180:-90:0.5cm) .. controls (0.4,-0.5) and (0.6,0) .. (0.6,0) .. controls (0.4,0.5) and (0,0.5) .. (0,0.5);
                    \draw[thick] (1.3,0) arc (0:90:0.3cm);
                    \draw[thick] (1.3,0) arc (0:-90:0.3cm);
                    \draw[thick] (0,-0.5) .. controls (0.4,-0.5) and (0.6,0) .. (0.6,0) .. controls (0.75,0.3) and (1,0.3) .. (1.,0.3);
                    \draw[thick] (0,0.5) .. controls (0.4,0.5) and (0.6,0) .. (0.6,0) .. controls (0.75,-0.3) and (1,-0.3) .. (1.,-0.3);
                \end{scope}
    
                \draw[very thick, ->, >=stealth] (-0.3,0.2) -- (-1.5,-1);
                \draw[very thick, ->, >=stealth] (1.1,0.2) -- (2.3,-1);

                \begin{scope}[yshift=-0.5cm]
                \begin{scope}[scale = 0.6, xshift = -4.2cm,yshift = -1.6cm]
                    \shade[outer color = purple!60!white, inner color = blue!10!white] (0,0) circle (0.1);
                    \shade[outer color = blue!60!white, inner color = blue!20!white] (1,0) circle (0.45);
    
                    \begin{scope}[rotate=180, xshift=-1cm]                    
                        \draw[thick] (0,0.5) arc (90:180:0.5cm) arc (-180:-90:0.5cm) .. controls (0.4,-0.5) and (0.6,0) .. (0.6,0) .. controls (0.4,0.5) and (0,0.5) .. (0,0.5);
                        \draw[thick] (1.3,0) arc (0:90:0.3cm);
                        \draw[thick] (1.3,0) arc (0:-90:0.3cm);
                        \draw[thick] (0,-0.5) .. controls (0.4,-0.5) and (0.6,0) .. (0.6,0) .. controls (0.75,0.3) and (1,0.3) .. (1.,0.3);
                        \draw[thick] (0,0.5) .. controls (0.4,0.5) and (0.6,0) .. (0.6,0) .. controls (0.75,-0.3) and (1,-0.3) .. (1.,-0.3);
                    \end{scope}
                \end{scope}
                
                \begin{scope}[scale = 0.6, xshift = 3.5cm,yshift = -1.6cm]
                    \shade[outer color = purple!60!white, inner color = blue!10!white] (0,0) circle (0.1);
                    \shade[outer color = blue!60!white!60!yellow, inner color = blue!20!white!60!yellow] (1,0) circle (0.15);
                    \begin{scope}[yshift=0.025cm]
                        \filldraw[color=blue!60!white!60!yellow!85!red] (1,-0.05) arc (-90:-75:0.6cm) -- (1.15,0) arc (90:80:0.6cm) arc (320:270:0.3325cm) arc (270:220:0.3325cm) arc (100:90:0.6cm) -- (0.85,-0.03) arc (-105:-90:0.6cm);
                    \end{scope}
    
                    \begin{scope}[scale = 0.95,rotate=180, xshift=-1.05cm]                    
                        \draw[thick] (0,0.5) arc (90:180:0.5cm) arc (-180:-90:0.5cm) .. controls (0.4,-0.5) and (0.6,0) .. (0.6,0) .. controls (0.4,0.5) and (0,0.5) .. (0,0.5);
                        
                    \end{scope}
                    \begin{scope}[scale = 0.7]
                        \draw[thick] (0,0.5) arc (90:180:0.5cm) arc (-180:-90:0.5cm) .. controls (0.4,-0.5) and (0.6,0) .. (0.6,0) .. controls (0.4,0.5) and (0,0.5) .. (0,0.5);
                    \end{scope}
                \end{scope}
    
                \begin{scope}[scale = 0.6, xshift = -4.2cm,yshift = -3.1cm]
                    \shade[outer color = black, inner color = black] (0,0) circle (0.05);
                    \shade[outer color = blue!60!white, inner color = blue!20!white] (1,0) circle (0.45);
    
                    \begin{scope}[rotate=180, xshift=-1cm]                    
                        \draw[thick] (0,0.5) arc (90:180:0.5cm) arc (-180:-90:0.5cm) .. controls (0.4,-0.5) and (0.6,0) .. (0.6,0) .. controls (0.4,0.5) and (0,0.5) .. (0,0.5);
                        \draw[thick] (1.3,0) arc (0:90:0.3cm);
                        \draw[thick] (1.3,0) arc (0:-90:0.3cm);
                        \draw[thick] (0,-0.5) .. controls (0.4,-0.5) and (0.6,0) .. (0.6,0) .. controls (0.75,0.3) and (1,0.3) .. (1.,0.3);
                        \draw[thick] (0,0.5) .. controls (0.4,0.5) and (0.6,0) .. (0.6,0) .. controls (0.75,-0.3) and (1,-0.3) .. (1.,-0.3);
                    \end{scope}
                \end{scope}

                \begin{scope}[scale = 0.6, xshift = 3.5cm,yshift = -3.1cm]
                    \shade[outer color = black, inner color = black] (0,0) circle (0.05);
                    \shade[outer color = blue!60!white!60!yellow, inner color = blue!20!white!60!yellow] (1,0) circle (0.15);
                    \begin{scope}[yshift=0.025cm]
                        \filldraw[color=blue!60!white!60!yellow!85!red] (1,-0.05) arc (-90:-75:0.6cm) -- (1.15,0) arc (90:80:0.6cm) arc (320:270:0.3325cm) arc (270:220:0.3325cm) arc (100:90:0.6cm) -- (0.85,-0.03) arc (-105:-90:0.6cm);
                    \end{scope}
    
                    \begin{scope}[scale = 0.95,rotate=180, xshift=-1.05cm]                    
                        \draw[thick] (0,0.5) arc (90:180:0.5cm) arc (-180:-90:0.5cm) .. controls (0.4,-0.5) and (0.6,0) .. (0.6,0) .. controls (0.4,0.5) and (0,0.5) .. (0,0.5);
                        
                    \end{scope}
                    \begin{scope}[scale = 0.7]
                        \draw[thick] (0,0.5) arc (90:180:0.5cm) arc (-180:-90:0.5cm) .. controls (0.4,-0.5) and (0.6,0) .. (0.6,0) .. controls (0.4,0.5) and (0,0.5) .. (0,0.5);
                    \end{scope}
                \end{scope}

                \begin{scope}[scale = 0.6, xshift = -4.2cm,yshift = -4.6cm]
                    \shade[outer color = black, inner color = black] (0,0) circle (0.05);
    
                    \begin{scope}[rotate=180, xshift=-1cm]          
                        \filldraw[color=blue!60!white] (0,0.5) arc (90:180:0.5cm) arc (-180:-90:0.5cm) .. controls (0.4,-0.5) and (0.6,0) .. (0.6,0) .. controls (0.4,0.5) and (0,0.5) .. (0,0.5);   
                        \shade[outer color = blue!60!white, inner color = blue!20!white] (0,0) circle (0.50);  
                        
                        \draw[thick] (0,0.5) arc (90:180:0.5cm) arc (-180:-90:0.5cm) .. controls (0.4,-0.5) and (0.6,0) .. (0.6,0) .. controls (0.4,0.5) and (0,0.5) .. (0,0.5);
                        \draw[thick] (1.3,0) arc (0:90:0.3cm);
                        \draw[thick] (1.3,0) arc (0:-90:0.3cm);
                        \draw[thick] (0,-0.5) .. controls (0.4,-0.5) and (0.6,0) .. (0.6,0) .. controls (0.75,0.3) and (1,0.3) .. (1.,0.3);
                        \draw[thick] (0,0.5) .. controls (0.4,0.5) and (0.6,0) .. (0.6,0) .. controls (0.75,-0.3) and (1,-0.3) .. (1.,-0.3);
                    \end{scope}
                \end{scope}

                \begin{scope}[scale = 0.6, xshift = 3.5cm,yshift = -4.6cm]
                    \shade[outer color = black, inner color = black] (0,0) circle (0.05);
    
                    \begin{scope}[scale = 0.95,rotate=180, xshift=-1.05cm]     
                        \filldraw[color=blue!20!white!60!yellow!80!red] (0,0.5) arc (90:180:0.5cm) arc (-180:-90:0.5cm) .. controls (0.4,-0.5) and (0.6,0) .. (0.6,0) .. controls (0.4,0.5) and (0,0.5) .. (0,0.5);   
                        \shade[outer color = blue!10!white!50!yellow!70!red, inner color = blue!20!white!60!yellow] (0,0) circle (0.5);
                        
                        \draw[thick] (0,0.5) arc (90:180:0.5cm) arc (-180:-90:0.5cm) .. controls (0.4,-0.5) and (0.6,0) .. (0.6,0) .. controls (0.4,0.5) and (0,0.5) .. (0,0.5);
                        
                    \end{scope}
                    \begin{scope}[scale = 0.7]
                        \draw[thick] (0,0.5) arc (90:180:0.5cm) arc (-180:-90:0.5cm) .. controls (0.4,-0.5) and (0.6,0) .. (0.6,0) .. controls (0.4,0.5) and (0,0.5) .. (0,0.5);
                    \end{scope}
                \end{scope}

                \begin{scope}[scale = 0.6, xshift = -4.2cm,yshift = -6.1cm]
                    \shade[outer color = black, inner color = black] (0,0) circle (0.05);
    
                    \begin{scope}[rotate=180, xshift=-1cm]          
                        \shade[outer color = purple!60!white!60!blue, inner color = purple!20!white] (0,0) circle (0.25);  
                        
                        \draw[thick] (0,0.5) arc (90:180:0.5cm) arc (-180:-90:0.5cm) .. controls (0.4,-0.5) and (0.6,0) .. (0.6,0) .. controls (0.4,0.5) and (0,0.5) .. (0,0.5);
                        \draw[thick] (1.3,0) arc (0:90:0.3cm);
                        \draw[thick] (1.3,0) arc (0:-90:0.3cm);
                        \draw[thick] (0,-0.5) .. controls (0.4,-0.5) and (0.6,0) .. (0.6,0) .. controls (0.75,0.3) and (1,0.3) .. (1.,0.3);
                        \draw[thick] (0,0.5) .. controls (0.4,0.5) and (0.6,0) .. (0.6,0) .. controls (0.75,-0.3) and (1,-0.3) .. (1.,-0.3);
                    \end{scope}
                \end{scope}

                \begin{scope}[scale = 0.6, xshift = 3.5cm,yshift = -6.1cm]
                    \shade[outer color = black, inner color = black] (0,0) circle (0.05);
    
                    \begin{scope}[scale = 0.95,rotate=180, xshift=-1.05cm]     
                        \shade[outer color = purple!60!white, inner color = purple!20!white] (0,0) circle (0.1);
                        
                        \draw[thick] (0,0.5) arc (90:180:0.5cm) arc (-180:-90:0.5cm) .. controls (0.4,-0.5) and (0.6,0) .. (0.6,0) .. controls (0.4,0.5) and (0,0.5) .. (0,0.5);
                        
                    \end{scope}
                    \begin{scope}[scale = 0.7]
                        \draw[thick] (0,0.5) arc (90:180:0.5cm) arc (-180:-90:0.5cm) .. controls (0.4,-0.5) and (0.6,0) .. (0.6,0) .. controls (0.4,0.5) and (0,0.5) .. (0,0.5);
                    \end{scope}
                \end{scope}
                \end{scope}
                
                \draw (-0.9,1.5) node {ZAMS};
                \draw (2.3,1.625) node {O-type + B-type star are born};
                \draw (2.3,1.375) node {and start core-H burning};
                
                \draw (-0.9,0.6) node {RLOF};
                \draw (2.25,0.6+0.25) node {primary expands and };
                \draw (2.25,0.6) node {fills its Roche lobe (RL)};
                \draw (2.25,0.6-0.25) node {$\Rightarrow$ mass transfer};

                \draw (-1.7,-0.3+0.125) node {conservative?};
                \draw (-1.7,-0.3-0.125) node {(short-period?)};
                \draw (2.7,-0.3+0.125) node {nonconservative?};
                \draw (2.7,-0.3-0.125) node {(long-period?)};

                \draw (-3.25,-1.45) node {WR+O};
                \draw (-0.7,-1.45+0.125) node {prim. = hotter WR};
                \draw (-0.7,-1.45-0.125) node {sec. rejuvenated};
                
                \draw (-3.25,-2.35) node {BH+O};
                \draw (-0.7,-2.35+0.125) node {prim. collapses};
                \draw (-0.7,-2.35-0.125) node {to black hole};

                \draw (-3.25,-3.25) node {2. RLOF};
                \draw (-0.7,-3.25+0.125) node {sec. expands to RL};
                \draw (-0.7,-3.25-0.125) node {$\Rightarrow$ mass transfer};
                
                \draw (-3.25,-4.15) node {BH+WR};
                \draw (-0.7,-4.15) node {sec. = cooler WR};

                \draw (1.325,-1.45) node {WR+B/Be};
                \draw (3.9,-1.45+0.125) node {prim. = hotter WR};
                \draw (3.9,-1.45-0.125) node {sec. rapid rotator};
                
                \draw (1.325,-2.35) node {BH+B/Be};
                \draw (3.9,-2.35+0.125) node {prim. = collapses};
                \draw (3.9,-2.35-0.125) node {to black hole};
                
                \draw (1.1,-3.25) node {BH+B-/Y-/RSG};
                \draw (3.9,-3.25+0.125) node {sec. expands to RL};
                \draw (3.9,-3.25-0.125) node {$\Rightarrow$ mass transfer};
                
                \draw (1.35,-4.15) node {BH+WR};
                \draw (3.9,-4.15) node {sec. = hotter WR};
                           
            \end{tikzpicture}
        \end{minipage}\hfill
        \begin{minipage}[c]{0.2\textwidth}
            \caption{Sketch of the evolutionary stages for possible formation channels of hotter and cooler WR stars under the assumption of stable mass transfer.}
            \rule{0ex}{26em}
            \label{fig:evolution}
        \end{minipage}
    \end{figure*}

        Figure~\ref{fig:evolution} depicts a simplified picture of the different formation channels of hotter and cooler WR stars. The flow chart is based on those models in our small grid that have a stable mass-transfer phase.
    
        The evolutionary tracks of the primaries with initial masses of $30\,\msun$ are similar to those of AzV\,14. All of the primary stars expand and initiate mass-transfer events, during which they lose most of the H-rich envelope, resulting in the formation of hotter WR stars (${\log T_\mathrm{eff}>4.9}$). The evolution of the secondaries depends on the initial parameters. The secondary in the system with initial period of $\SI{5}{d}$ accretes about $\sim5\,\msun$ of material, leading to a rejuvenation of the core. After hydrogen is depleted in its core, the stellar model expands and quickly initiates mass transfer, stripping off parts of the accreted envelope. After the mass-transfer event the secondary has a temperature of ${\log T_\mathrm{eff}\approx4.7}$ and a high surface H-abundance of $X_\mathrm{H}=0.5$ (left side of Fig.~\ref{fig:evolution}). On the other hand, the secondary in the system with initial period of $\SI{50}{d}$ accretes less than $0.5\,\msun$. This is a big difference to the short-period model, which is explained by the fact that in our models accretion is only allowed when the accretor can avoid rapid rotation. In the short-period binaries, the stars are tidally locked, slowing down the rotation sufficiently for the accretor to stay below critical rotation during the accretion process. In a long-period binary, the star spins up quickly, limiting the amount of mass that can be accreted \citep{pet1:05,dem1:07,yon1:16}. In such long-period systems (right side of Fig.~\ref{fig:evolution}) the secondary initiates mass transfer after core-H burning during its evolution toward the blue (BSG) and yellow supergiant (YSG) phase, leading to the formation of a hotter WR star with ${\log T_\mathrm{eff}\approx4.9}$ (i.e., the same temperature regime that is also populated by the primary models). In even longer period systems ($P=\SI{500}{d}$) mass transfer is initiates when the star is evolving toward a red supergiant (RSG). The model has already formed a large convective envelope which expands adiabatically, making mass-transfer unstable. Potentially, this leads to  a common envelope evolution which is not modeled here.

        The models with initial primary masses of $45\,\msun$ and initial periods of $5$ and $\SI{50}{d}$ also predict the formation of hotter WR stars (${\log T_\mathrm{eff}>4.85}$). However, the primary model with initial period of $\SI{500}{d}$ is on its way to becoming a RSG and has already formed a large outer convection region. Similar to the models presented above, this leads to an unstable mass transfer and possibly a common envelope evolution. In the model with an initial period of $\SI{5}{d}$ the secondary accretes a significant amount of mass (${>10\,\msun}$) and initiates mass transfer after its main sequence evolution. After the mass transfer event the surface H fraction drops to $X_\mathrm{H}=0.35$ and the temperature is ${\log T_\mathrm{eff}>4.7}$. The secondary will end its life as a cooler WR (left side of Fig.~\ref{fig:evolution}). For an initially wider orbit ($\SI{50}{d}$) the secondary  accretes only negligible amounts of material and, similar to the primary models, evolves into a hotter WR star (right side of Fig.~\ref{fig:evolution}). 
    
        The primary models with initial masses of $60\,\msun$ populate a wide range of effective temperatures (${{\log T_\mathrm{eff}=\numrange{4.65}{5.0}}}$). The broad temperature range can be explained by the changing efficiency of envelope stripping and its dependence on initial period and mass ratio. The efficiency is linked to the point in the primary's evolution when it transitions into the core-He burning stage, which happens before or during the mass transfer \citep[see also][their figure 3]{kle1:22}. In the primary model of the system with initial period of $\SI{5}{d}$ the mass transfer is very efficient in removing the H-rich envelope and with the help of the strong WR winds the model is able to remove a significant amount of the envelope, making the star appear hot. On the other hand, in the primary model with initial period of $\SI{500}{d}$ the mass-transfer is less efficient, making the WR star appear cooler. The secondary in the system with an initial period of $\SI{5}{d}$ accretes a significant amount of mass ($>10\,\msun$). It initiates mass transfer after the main sequence, stripping off parts of the accreted envelope, resulting in the formation of a cooler WR. The secondary in the system with an initial period of $\SI{50}{d}$ behaves similarly to its corresponding primary model and forms a hotter WR star. The secondary in the system with initial period $\SI{500}{d}$ initiates mass transfer when it is on its way to becoming a RSG and has already formed an extended outer convection zone. Mass-transfer in this model is unstable.

    \subsection{Bimodal temperature distribution of WR stars at low metallicity}
    \label{sec:pred_WR}
    
     	\begin{figure}[tbhp]
     	    \centering
     	    \includegraphics[trim= 0.5cm 0.5cm 0.7cm 1.8cm ,clip ,width=0.5\textwidth]{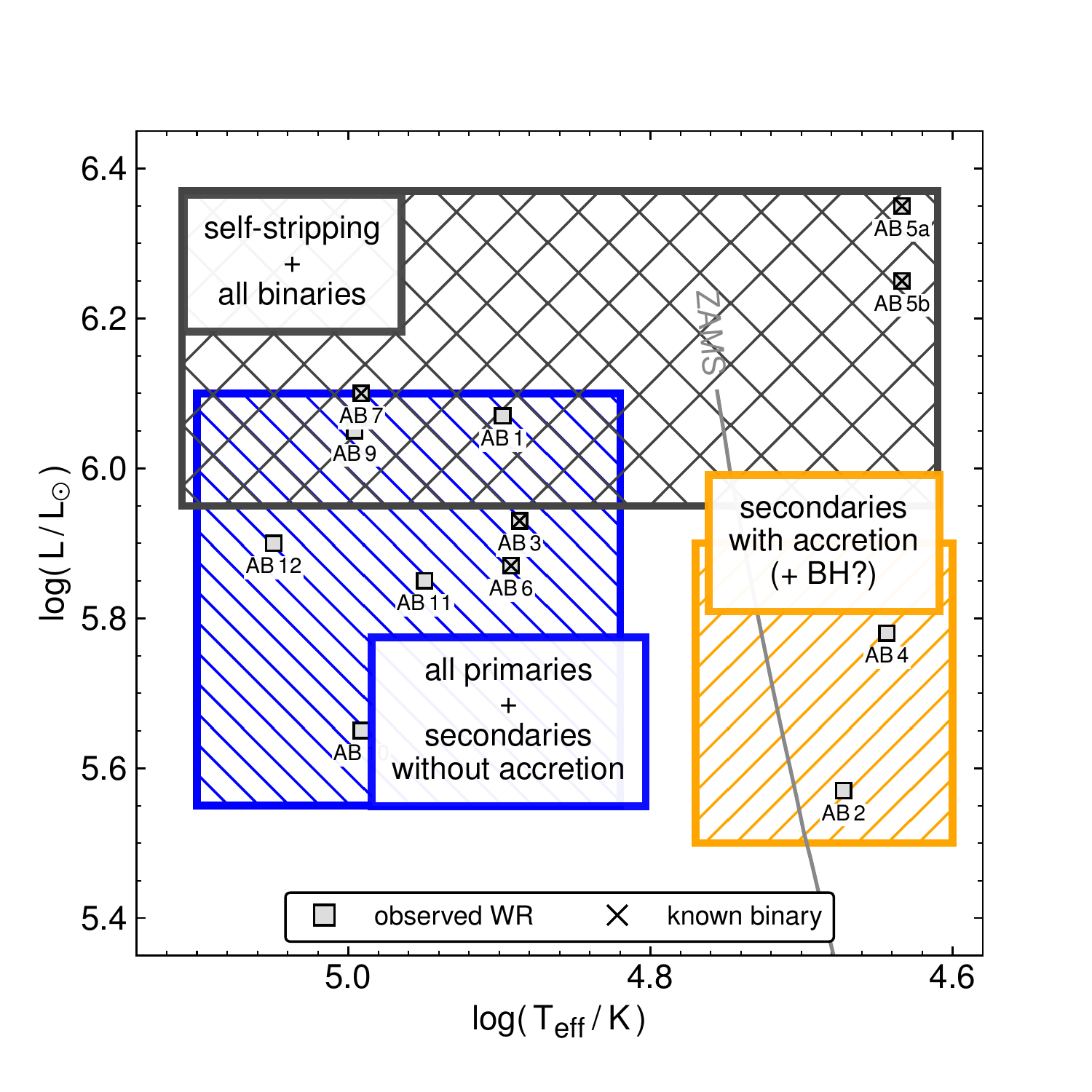}
     	    \caption{HRD containing the positions of the observed WR stars in the SMC, over-plotted by a sketch stating the different evolutionary channels of single stars and binaries undergoing stable mass transfer, leading to the different marked regions.}
     	    \label{fig:sketch}
     	\end{figure}

        From considering different evolutionary pathways probed by our exploratory model grid, we learned that primaries and secondaries which have not accreted a significant amount of material evolve to hotter WR stars.  Only secondaries that have accreted a fair amount of mass (i.e., those with the shortest periods) evolve into cooler WR stars. There are two main factors that may be responsible for this behavior. Firstly, evolved accretors are characterized by a steep chemical gradient at the core-envelope boundary, producing cooler WR stars once the outer envelope is lost \citep[][their figure 9]{sch1:18}. Secondly, accretors that are not fully rejuvenated tend to remain compact after the end of MS and begin the core-He burning phase as BSGs \citep{van1:13,jus1:14}. Mass transfer from such stars leads to less efficient envelope stripping and cooler WR products \citep{kle1:22}. This is different from the single star models \citep[e.g.,][]{geo1:12,cho1:16,eld1:17,lim1:18} which can only explain the hottest and most luminous WR stars.
        
        Based on our models, we hypothesize that there must be a bimodal temperature distribution of faint WR stars at low metallicity. First, hotter WR stars of any luminosity can be well explained by the primary models as well as by the secondary models that did accrete negligible amounts of material (i.e., those in wide orbits). Second, cool and faint WR stars (${{\log T_\mathrm{eff}=\numrange{4.65}{4.7}}}$ and ${\log (L/\lsun)\simeq \numrange{5.5}{5.9}}$) must arise from secondaries that have accreted a significant amount of material, leading to a rejuvenation and high envelope-to-core mass ratios. Third, hotter and luminous WR stars (${{\log T_\mathrm{eff}=\numrange{4.65}{4.7}}}$ and ${\log (L/\lsun)\simeq \numrange{5.9}{6.2}}$) can be explained by primary, secondary, and single star models. We note that according to our models the cool and faint WR stars should all be accompanied by a compact companion, which can avoid detection and can help to explain observed apparently single cooler WR stars. We sketched the morphology of the WR population in the SMC in Fig.~\ref{fig:sketch}.

        It is less clear whether cooler WR stars at higher metallicities are formed in the same way as in low-metallcity galaxies. Two reasons should be considered. First, at high metallicity, the observed effective temperatures of WR stars can be lower, due to stronger winds \citep[e.g.,][]{san1:20,san1:23}, also termed ``dynamic inflation'' \citep{gra1:18}. Second, at higher metallicity WR stars additionally suffer from the effect of hydrostatically inflated envelopes \citep[e.g.,][]{gra1:12,san1:15}, yielding again lower effective temperatures. Third, the stellar winds in the pre-WR stage at higher metallicity are stronger, stripping off more of the H-rich envelope, which is left after mass transfer. While the latter leads to the formation of hotter WR stars (as shown in Sect.~\ref{sec:enhanced_mdot}), the effect on the observed temperature might partially be counter-balanced by the higher wind densities. All these effects make it difficult to distinguish between WR stars formed from stars which have accreted significant amounts of material in the past and intrinsically inflated WR stars.
        
        Hence, in order to clearly see a bimodal temperature distribution of WR stars originating from post interaction binaries, populations at low metallcity must be considered. However, the number of WR stars decreases with decreasing metallicity \citep[e.g.,][]{she1:20}, enforcing us to rely on small number statistics. In the SMC, the sample of WR stars is complete which minimizes selection biases. Therefore, despite the small number statistics, it remains the best representative sample of an WR population at low metallicity. In order to confirm or falsify our predictions, further observations of complete populations of WR stars in other low metallicity galaxies in combination with calculations of binary evolutionary models are necessary.

        Our findings and other recent results \citep{ren1:21,ren1:22} indicate that evolution of past accretors may be systematically different to those of normal (primary) stars.  This impacts our understanding of stellar evolution and feedback. For instance, in the SMC hotter WR stars like AB\,10 ($\log(Q_\HeII{}/\mathrm{s^{-1}})=48.1$), have strong \HeII{} ionizing fluxes, while cooler WR stars like AB\,4 ($\log(Q_\HeII{}/\mathrm{s^{-1}})=37.5$) have five orders of magnitude lower ionizing flux. Broader implications for binary evolution channels, proposed in this work, are yet to be explored.

    \subsection{Observational fingerprints of previous evolutionary channels}

        Our model predicts that the observed effective temperature is one of the key diagnostics to differentiate between the evolutionary origin of WN type stars at low metallicity, as explained in Sect.~\ref{sec:pred_WR}. However, for WN stars originating from past accretors, a low effective temperature is not the only diagnostic.
    
        The surface H-abundance can be considered as another clue, as all of our rejuvenated secondary model show ${X_\mathrm{H}\gtrsim0.3}$, while for their corresponding primary models the surface H-abundance are noticeably lower ${X_\mathrm{H}\sim0.2}$.  However, for our most luminous models (${M_\mathrm{ini,\,1}=60\,\msun}$) in the systems with the widest orbits, the primaries have low temperatures and the H-abundances are comparable to those of the rejuvenated secondaries. Hence, a high hydrogen abundance and a low temperature  are not a robust criterion to identify the previous evolutionary path.
    
        In our search for additional fingerprints to detect past accretors, we identified a difference in the predicted CNO surface composition of normal (primary) stars and the rejuvenated secondaries. It is expected that both, the primaries and nonrejuvenated secondaries, have CNO equilibrium composition (${X_\mathrm{C}=\numrange{1}{2e-5}}$, ${X_\mathrm{N}=\num{139e-5}}$, and ${X_\mathrm{O}=\numrange{1}{3e-5}}$).  In a direct comparison, the models of the rejuvenated secondaries have a decreased surface N-abundance, while the surface O-abundance is increased by about one order of magnitude (${X_\mathrm{C}=\numrange{1}{2e-5}}$, ${X_\mathrm{N}\approx\num{120e-5}}$, and ${X_\mathrm{O}=\numrange{10}{25e-5}}$), due to the formation of a shallow chemical gradient extending through the accreted envelope. Therefore, we predict that oxygen lines should be present in spectra of cooler WN stars formed from past accretors.
        
        For the analysis of larger WN samples \citep[e.g.,][]{ham1:06,san1:14,hai1:15,she1:16}, traditionally the N-abundance has been fixed to a value reflecting CNO equilibrium, while oxygen has not been included at all, due to its weak or absent spectral imprint in the optical regime. To predict the spectral appearance of WN stars which originate from stars that have accreted significant amounts of material in the past, synthetic spectra now need to include oxygen as well. Given the temperatures of WR stars \OIV{} and \OV{} are expected to be dominant ions.
        
\begin{figure}[tbp]
        \centering
        \includegraphics[trim= 0cm 0cm 0cm 0cm ,clip ,width=0.4\textwidth]{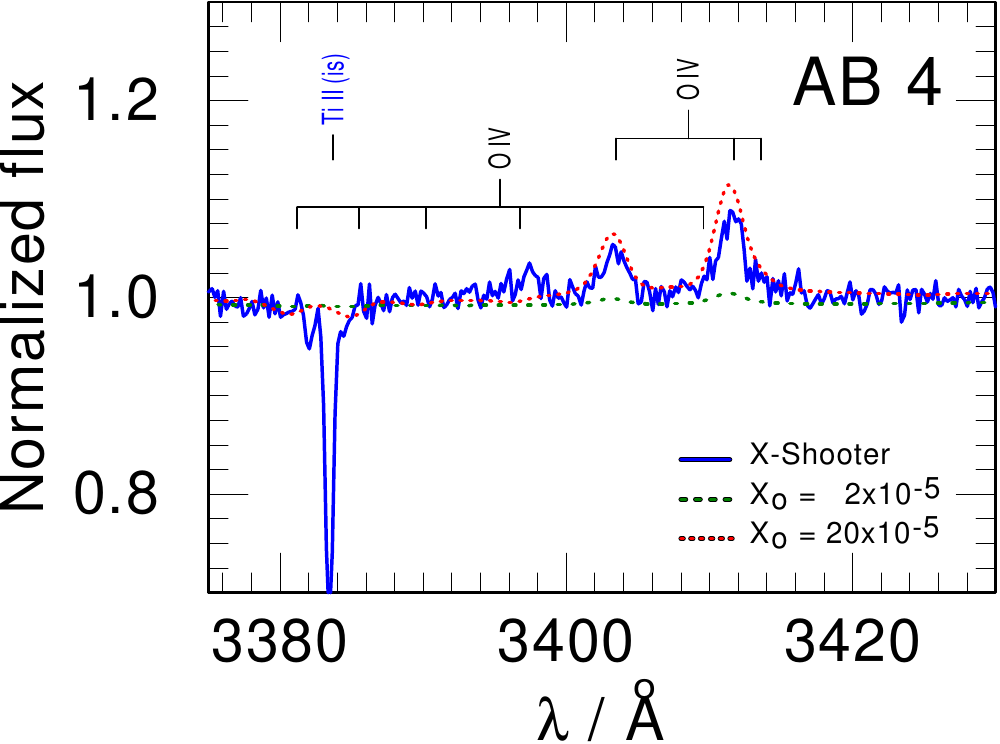}
        \caption{Close-up view of the \OIV{} multiplet in the X-Shooter spectrum of the WR star SMC AB~4. The observed spectrum is shown as a blue solid line, the synthetic spectra are calculated with an oxygen abundance of $X_\mathrm{O}=\num{2e-5}$ and $X_\mathrm{O}=\num{20e-5}$ and are shown as green dotted and red dotted lines, respectively.}
        \label{fig:AB4_OIV}
    \end{figure}

        In order to check this prediction, we select the WR star SMC\,AB\,4 (WN6h), as its empirical position in the HRD can only be explained by our models of secondaries that have accreted material during mass transfer. A spectrum of AB4 taken with X-Shooter at the ESO-VLT within the ESO program 106.211Z (PI J. S. Vink) is available from the archive. We re-calculated the stellar atmosphere model of AB\,4, from \citet{hai1:15} with an enhanced O-abundances of $X_\mathrm{O}=\num{20e-5}$ as well as with the CNO equilibrium value of $X_\mathrm{O}=\num{2e-5}$. Indeed, the model with ten times enhanced oxygen abundance reproduces the \OIV{} lines at wavelengths $\SI{3403}{\AA}$ and $\SI{3414}{\AA}$ just as observed, while the CNO equilibrium abundance fails by far (see Fig.~\ref{fig:AB4_OIV}).
        
        As illustrated by this example, WN stars with enhanced oxygen abundance, such as AB\,4, may be explained by the evolutionary models assuming that they accreted a significant amount of matter in the past. We suggest that the oxygen abundance of WN type stars is an important diagnostic which can help to identify WR stars originating from binary components that have accreted significant amounts of material in the past.    
        
    \subsection{Sensitivity of the evolutionary model to different parameters}
    \label{sec:sensitivity}

        Estimating uncertainties in stellar evolution models is a nontrivial task, as it does not only depend on the numerical uncertainty but also on the physical assumptions made within the model itself. Stellar evolution models rely on many free parameters that can alter the evolution of a star, for instance mass-loss and mixing processes. Exploring the impact of all free parameters, quickly opens a large parameter space and needs intensive computational power. Such an investigation is far beyond the scope of the present paper. In the following, we briefly discuss two examples of how these parameters can impact our predictions on the temperature distribution of WR stars in low-metallicity environments.
        
    \subsubsection{Example 1: Enhanced mass-loss rates}    
    \label{sec:enhanced_mdot}
    
        \begin{figure*}[t]
     	    \centering
     	    \includegraphics[trim= 1.5cm 1.cm 1.5cm 0cm ,clip ,width=\textwidth]{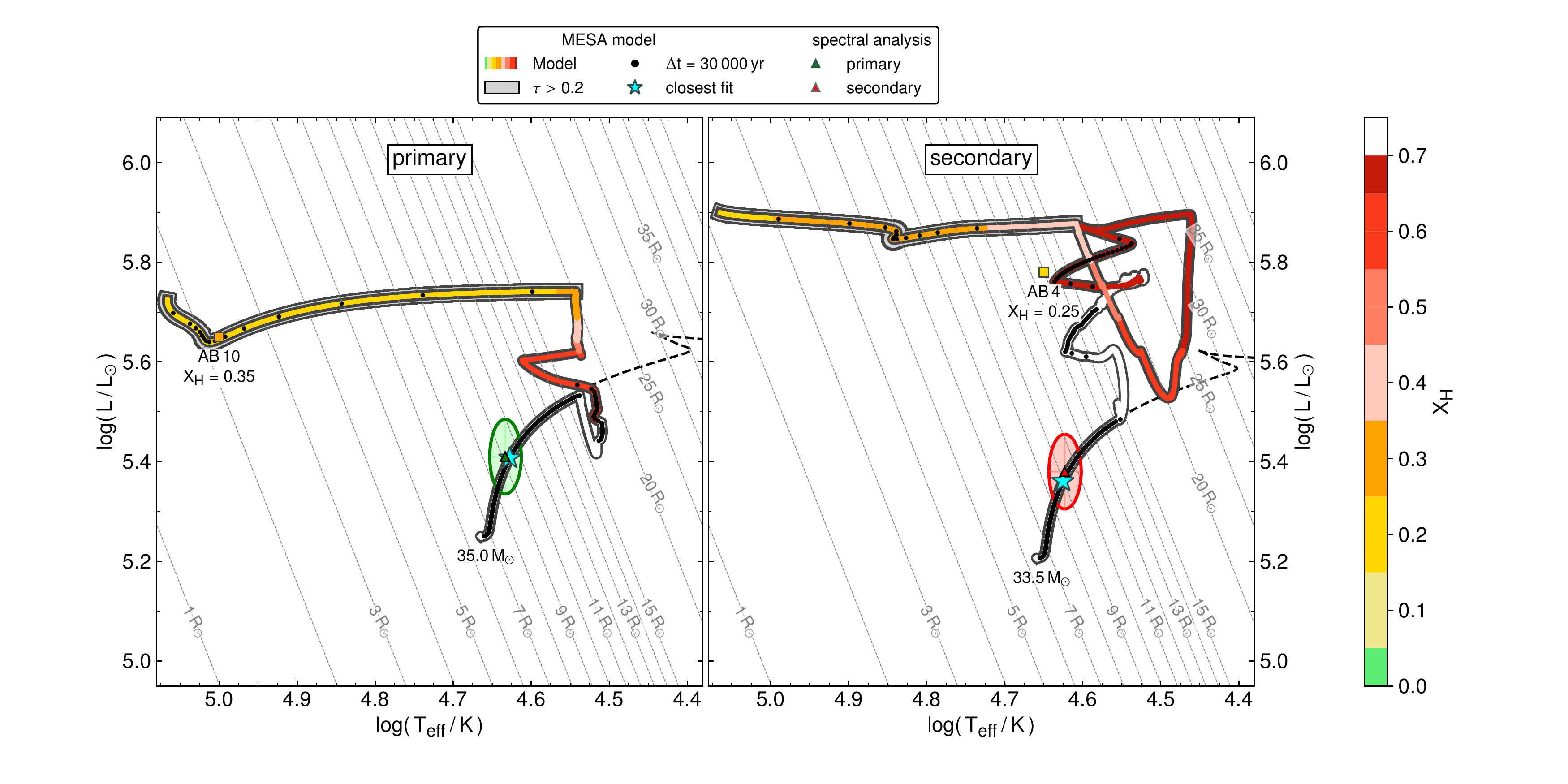}
     	    \caption{Same as Fig.~\ref{fig:mesa_best_fit}, but now with the WR mass-loss rates enhanced by a factor $3$.}
     	    \label{fig:3xshenar}
     	\end{figure*}
        
        In the binary model presented in this paper, the mass-transfer takes place while the primary is still core-H burning. Therefore, the evolution of the secondary is not really sensitive to the mass-loss rates of O stars. However, both stars spend a significant amount of their lifetime in the WR stage. Hence, we decided to test how the evolution would be affected if the WR mass-loss rate is increased by a factor of 3.
        
        Figure~\ref{fig:3xshenar} shows the resulting tracks. By comparison to Fig.~\ref{fig:mesa_best_fit}, the predicted temperature regimes and the surface chemical composition during the WR phases changes. The primary loses a larger fraction of its H-poor envelope and spends more time at higher temperatures. At the end of core-He burning the stellar evolution model starts to contract instead of expanding. On the other hand, the stellar evolution model of the secondary shows even stronger response to the increased mass-loss rate. Instead of becoming a cooler WR star with plenty of hydrogen in the envelope, it is now able to remove large fractions of this envelope and position itself at higher temperatures in the HRD. We note that the region in the HRD where the stellar evolution model of the secondary is located is not populated by an observational counterpart.
 	        
        The above example shows  that the choice of the mass-loss recipe during different evolutionary phases can have a drastic impact on the evolution of massive stars. Only when using adequate mass-loss recipes, one can explain the cooler WR stars observed in the SMC. We note that in the case of WR winds being weaker than we assume in this work, the prediction on the morphology of the temperature distribution of WR stars would shift to lower metallicities.
        
    \subsubsection{Example 2: More efficient mixing}
    
     	\begin{figure*}[thpb]
     	    \centering
     	    \includegraphics[trim= 1.5cm 1.cm 1.5cm 0cm ,clip ,width=\textwidth]{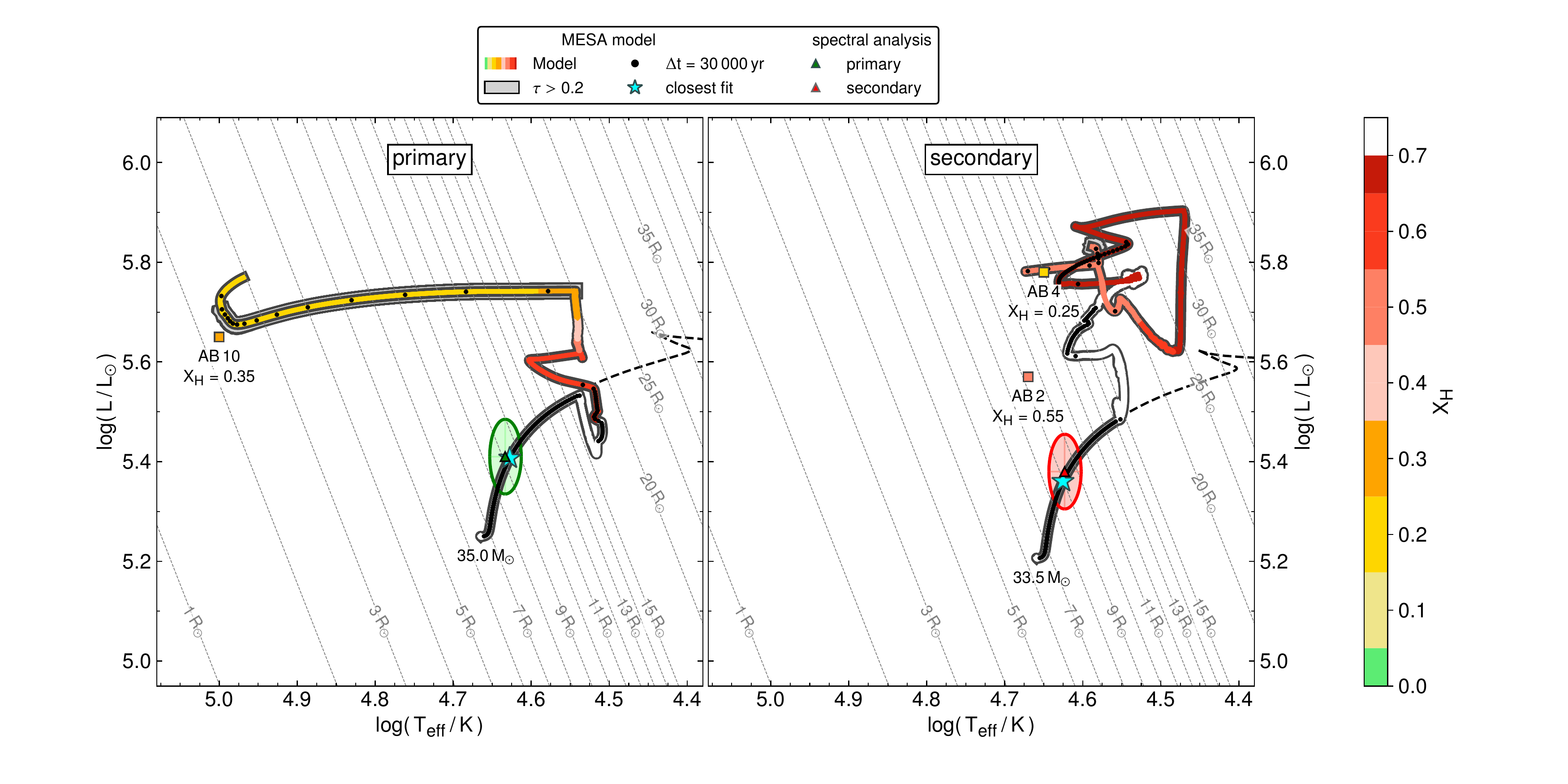}
     	    \caption{Same as Fig.~\ref{fig:mesa_best_fit}, but now with an increased efficiency of semiconvection.}
     	    \label{fig:sc10}
     	\end{figure*}
      
        There are several mixing processes within stellar interiors, including convective mixing, semiconvective mixing, overshooting, rotational mixing, pulsational mixing, etc. Typically, these mixing processes are parameterized by free parameters, for example the mean free path a photon can travel, or the efficiency of a specific mixing process which cannot be predicted by theory. In the literature there are many works trying to limit the parameter space of the different free parameters \citep[e.g.,][]{sch1:19,hig1:19,gil1:21,mic1:21}. In this section, we want to showcase how changing one of the free parameters can impact the evolution of a star. 
        
        One of the least constrained , yet important, mixing process is the efficiency of semiconvective mixing. \citet{sch1:19} used the population of the BSGs and RGSs to constrain semiconvective mixing efficiencies, but were able only to limit it to $\alpha_\mathrm{sc}\gtrsim1$. They report that higher values only barely impact the number ratio of blue and red supergiants in galaxies. In our models presented above we used $\alpha_\mathrm{sc}=1$. For the models presented in this section, however, we increased the efficiency to $\alpha_\mathrm{sc}=10$ in order to see how it changes our understanding of the evolution of the binary.

        The evolutionary tracks of the primary and secondary with the more efficient semiconvective mixing are shown in Fig.~\ref{fig:sc10}.  By comparing these tracks to those shown in Fig.~\ref{fig:mesa_best_fit}, one can see that the  evolution of the primary barely changes. Semiconvective mixing becomes important only in evolutionary stages after the main sequence. The primary begins  mass transfer already during its main sequence evolution and  spends its post-main sequence evolution as a WR star, leaving the semiconvective mixing no time to impact the evolution. On the other hand, the evolution of the secondary changes because during the accretion phase it forms extensive convective and semiconvective regions in its envelope. The secondary initiates mass-transfer after the main sequence evolution, giving semiconvection some time to efficiently mix material. In addition, due to more efficient semiconvection, the secondary expands less in the transition to the core-He burning phase and initiates mass transfer at a somewhat more advanced evolutionary stage, consequently losing less mass during mass-transfer and producing a cooler WR star with an even higher surface H abundance. This is reflected in the higher surface H-abundance and the cooler temperature. \citet{kle1:22} recently reported something similar: Binary models at low metallicity with highly efficient semiconvective mixing will not become WR/helium stars after Case~B mass transfer. Their models managed to stay at the position in the HRD at which they initiated the mass transfer and hence are core-He burning stars, disguised as O- or B-type stars. 
        It is worth mentioning that  according to our criterion based on the optical depth, the secondary of the binary evolutionary model with efficient semiconvective mixing would spend only a short time as a WR star at low temperature. 
        
        These two examples illustrate that binary evolution models are quite sensitive to the input parameters. A more detailed study based on grids of detailed binary evolutionary models and population synthesis are needed to confirm or disprove our predictions on the morphology of WR populations at low metallicity.

    \section{Summary and conclusions}
    \label{sec:conclusions}

        A consistent analysis of multi-epoch optical and UV spectra of  one of the earlierst O-type stars in the SMC, AzV\,14, reveals its binary nature. Furthermore, our analysis uncovered that the systems' two  components are very similar. The primary and secondary have temperatures of ${T_\mathrm{eff,\,1}=\SI{42.8\pm2}{kK}}$ and ${T_\mathrm{eff,\,2}=\SI{41.8\pm2}{kK}}$, luminosities of ${\log(L_1/\lsun)=5.41\pm0.15}$ and  ${\log(L_2/\lsun)=5.38\pm0.15}$ and surface gravities of ${\log(g_{1,2}/(\mathrm{cm\,s^{-2}})) = 4.0\pm0.2}$, respectively. The analysis of a TESS light curve confirms the binary nature of AzV\,14 with an orbital period of $\SI{3.7058\pm0.0013}{d}$. Their spectroscopic masses of ${M_1=32\pm8\,\msun}$ and ${M_2=31\pm8\,\msun}$ are confirmed by the orbital analysis. Both binary components drive weak stellar winds with mass-loss rates of about ${\log(\dot{M}/(\msunpyr))=-7.7\pm0.2}$. The relatively high observed X-ray emission of AzV\,14 is attributed to colliding winds.  

        The new empirically derived stellar parameters and the current orbital period  of AzV\,14 are well explained by our evolutionary models. In particular, the spectroscopic and orbital masses of the binary components are in agreement with the evolutionary ones. According to the evolutionary models, the components of AzV\,14 did not yet exchange mass. Most interestingly, the binary evolutionary model of AzV\,14 predicts that the primary will evolve into a hotter WR star, while the secondary is destined to evolve into a cooler WR star.

        Inspired by these results we calculated a small evolutionary model grid to investigate the conditions for the formation of hotter and cooler WR stars. Guided by these calculations, we anticipate that WR populations in low metallicity galaxies should show bimodal temperature distributions. According to our models, stars born in long-period binaries both evolve to hotter WR stars. On the other hand, for stars born in short-period binaries only the primary is destined to become a hot WR star, while the secondary evolves into a cooler WR star. In our models we assume that accretion in short-period binaries is efficient, leading to a rejuvenation of the core and a different envelope-to-core mass ratio. This eventually allows the past accretors to stay at lower temperatures during their WR stage.

        These results are sensitive to the input physics, but if turned out to be correct they would have a vast impact on our understanding of binary evolution at low metallicity, stellar feedback, and, hence, on galaxy formation and evolution. To test if our models are reliable and cooler WR stars originate from past accretors we introduced an additional criterion: the surface oxygen abundance. From standard evolutionary models one expects that a WR star has a surface composition which corresponds to the CNO equilibrium value. However, in our models the cooler WR stars, formed by stars that accreted a significant amount of material in the past, all have surface oxygen abundance which are increased by a factor of about ten. We tested this prediction with archival spectra of the apparently single WN star AB4 in the SMC. We detect oxygen lines in its optical spectrum, and measure the high oxygen abundance which is in agreement with our evolutionary model predictions. This empirically supports the new evolutionary pathway to the formation of WR stars at low metallcity proposed in this paper.
         	    
    \begin{acknowledgements}
        The authors are appreciative to the reviewer for useful comments which helped to improve the paper, and for their suggestions. The results presented in this paper are based on observations obtained with the NASA/ESA Hubble Space  Telescope, retrieved from MAST at the Space Telescope Science Institute (STScI). STScI is operated by the Association	of Universities for Research in Astronomy, Inc. under NASA contract NAS 5-26555. Support to MAST for these data are provided by the NASA Office of Space Science via grant NAG5-7584 and by other grants and contracts. The TESS data presented in this paper were obtained from MAST at the STScI. Funding for the TESS mission was provided by the NASA Explorer Program. Furthermore, its conclusions are based on observations collected at the European Southern Observatory (ESO) under the program 098.A-0049. The authors thank  the managing committee of XShootU and Andrea Mehner for preparing the OBs of the XShootU project. This work has made use of data from the European Space Agency (ESA) mission {\it Gaia} (\url{https://www.cosmos.esa.int/gaia}), processed by the {\it Gaia} Data Processing and Analysis Consortium (DPAC, \url{https://www.cosmos.esa.int/web/gaia/dpac/consortium}). Funding for the DPAC has been provided by national institutions, in particular the institutions participating in the {\it Gaia} Multilateral Agreement. DP and SRS acknowledge financial support by the Deutsches Zentrum f\"ur Luft und Raumfahrt (DLR) grants FKZ 50OR2005 and 50OR2108. AACS and VR acknowledge support by the Deutsche Forschungsgemeinschaft (DFG, German Research Foundation) in the form of an Emmy Noether Research Group -- Project-ID 445674056 (SA4064/1-1, PI Sander). DMB gratefully acknowledges a senior postdoctoral fellowship from the Research Foundation Flanders (FWO) with grant agreement number 1286521N. RK acknowledges financial support via the Heisenberg Research Grant funded by the German Research Foundation (DFG) under grant no.~KU 2849/9. CK acknowledges financial support from the Spanish Ministerio de Economía y Competitividad under grants AYA2016-79724-C4-4-P and PID2019-107408GB-C44, from Junta de Andalucía Excellence Project P18-FR-2664, and from the State Agency for Research of the Spanish MCIU through the ‘Center of Excellence Severo Ochoa’ award for the Instituto de Astrofísica de Andalucía (SEV-2017-0709).  TS acknowledges support from the European Union's Horizon 2020 under the Marie Skłodowska-Curie grant agreement No 101024605. The collaboration of coauthors was facilitated by support from the International Space Science Institute (ISSI, Bern).
        The authors are grateful to the Lorentz Center (Leiden) for organizing a stimulating workshop. We thank Paul Crowther for useful discussions as well as advice regarding the UVES spectrum and for providing the updated IR photometry.

    \end{acknowledgements}

 	\bibliographystyle{aa}                                                         
 	\bibliography{astro}

 \clearpage

    \begin{appendix}
 		\section{Spectroscopic and photometric data}
 		\label{app:spectra}
 		    
 		    The information on all observed spectra used in this work is given in Table~\ref{tab:spectra}. All spectra are in a reduced state from the individual observatories pipeline. Apart from the UVES spectrum,  all other spectra analyzed in this work had been  corrected for barycentric motion by the automated extraction pipeline. For the UVES spectrum we had to calculate this correction ourselves. With the help of the tool from \citet{wri1:14} we obtained ${\varv_\mathrm{bary}=\SI{-10.6}{km\,s^{-1}}}$. A summary of the estimated RVs and the calculated orbital phases of the individual observed spectra that are used in our analysis are listed in Table~\ref{tab:spectra2}. The photometric data employed for the SED fitting are listed in Table~\ref{tab:photometry}. 
            
            \begin{table*}[tbhp]
     		\centering
     		\small
     		\caption{Spectroscopic observations of AzV\,14.}
         		\begin{tabular}{cccccccc}\hline \hline \rule{0cm}{2.2ex}%
         			Instrument & Wavelength & Resolving power & Exposure time &S/N &  MJD$^{(a)}$ & \multicolumn{1}{c}{Program ID} & \multicolumn{1}{c}{PI}\\
         			& [\AA] & & [s] & & [d]  &\\
         			\hline \rule{0cm}{2.4ex}%
         			\rule{0cm}{2.2ex}FUSE & $\SIrange{950}{1\,150}{}$ & $20\,000$ & $6770$ & 7.4 & 51726.5778 & p1175301000 & J. B. Hutching\\
         			\rule{0cm}{2.2ex}HST/FOS & $\SIrange{1\,140}{1\,606}{}$ & 250 & 480 & $\approx20$ &  49985.5561 & Y25U0101T &  C. Robert\\
         			\rule{0cm}{2.2ex}HST/STIS & $\SIrange{1\,140}{1\,740}{}$ & $45\,800$ & 2767& $\approx66$  &  58981.1081 & 15629 & L. Mahy\\
         			\rule{0cm}{2.2ex}UVES & $\SIrange{3\,731}{4\,999}{}$ & $\approx41\,000$ & 2400& 47.3  &  52180.0263 & 67.D-0238(A) & P. Crowther \\
         			\rule{0cm}{2.2ex}X-Shooter~(2020) & $\SIrange{3\,000}{24\,800}{}$ & $\gtrsim5500$ & $\approx1000$ & $\gtrsim100$ &  59156.1955 & 106.211Z & J. S. Vink\\
         			\rule{0cm}{2.2ex}X-Shooter~(2022)& $\SIrange{3\,000}{24\,800}{}$ & $
         			\gtrsim5500$ & $900$ & $\gtrsim90$ &  59847.3603 & 109.22V0.001 & D. Pauli\vspace{0.3ex}\\
         			\hline
         		\end{tabular}
         		\label{tab:spectra}
                \rule{0cm}{2.8ex}%
                \begin{minipage}{0.95\linewidth}
                    \ignorespaces 
                    $^{(a)}$ Mid-exposure time in $\mathrm{HJD}-2400000.5$. 
                \end{minipage}
         	\end{table*}

            \begin{table}[tbhp]
     		\centering
     		\small
     		\caption{Fitted RV shifts  of the binary components of AzV\,14 in each spectrum. The given RVs are in addition to the RV of NGC\,261 (${\SI{148}{km\,s^{-1}}}$).  The phases are calculated according to our ephemeris (see Sect.~\ref{sec:phoebe}).}
         		\begin{tabular}{ccrr}\hline \hline \rule{0cm}{2.2ex}%
         			Instrument & Phase $\Phi^{(b)}$ & \multicolumn{1}{c}{$\mathrm{RV}_1$} & \multicolumn{1}{c}{$\mathrm{RV}_2$}\\
         			&  & \multicolumn{1}{c}{[$\si{km\,s^{-1}}$]} & \multicolumn{1}{c}{[$\si{km\,s^{-1}}$]}\\
         			\hline \rule{0cm}{2.4ex}%
         			\rule{0cm}{2.2ex}FUSE & 0.4021 & $62.7\pm4.7$ & $-95.3\pm4.3$ \\
         			\rule{0cm}{2.2ex}HST/FOS & 0.1924 & $92.9\pm6.2$ & $-113\pm6.4$ \\
         			\rule{0cm}{2.2ex}HST/STIS & -0.3173 & $-128.6\pm2.5$ & $113.2\pm2.8$ \\
         			\rule{0cm}{2.2ex}UVES & -0.1315 & $-101.9\pm2.6$ & $80.5\pm3.2$ \\
         			\rule{0cm}{2.2ex}X-Shooter~(2020) & -0.0302 & $-46.1\pm5.8$ & $22.3\pm5.9$ \\
         			\rule{0cm}{2.2ex}X-Shooter~(2022) & -0.3625 & $-112.4\pm2.6$ & $98.6\pm2.9$ \\
         			\hline
         		\end{tabular}
         		\label{tab:spectra2}
                \rule{0cm}{2.8ex}%
         	\end{table}  
         	
            \begin{table}[tbhp]
     		\centering
     		\caption{Photometry of AzV\,14. The UBI photometry from \citet{bon1:10}, V-band photometry from \citet{mey1:93}, JHK photometry from the 2MASS catalog \citep{cut1:03}, Gaia EDR3 G-band photometry \citep{gai1:16,gai1:21}, and YJ$_\mathrm{s}$K$_\mathrm{s}$ photometry from the VMC survey \citep{cio1:11}.}
         		\begin{tabular}{cc}\hline \hline \rule{0cm}{2.2ex}%
         			Band & Apparent magnitude\\
         			& [mag]\\
         			\hline \rule{0cm}{2.2ex}%
         			U & $12.53\pm0.03$ \\
         			B & $13.59\pm0.04$ \\
         			V & $13.7$ \\
         			G & $13.822\pm0.003$ \\
         			G$_\mathrm{BP}$ & $13.675\pm0.003$ \\
         			G$_\mathrm{RP}$ & $13.964\pm0.004$ \\
         			I & $13.85\pm0.04$ \\
         			J & $14.193\pm0.035$ \\
         			H & $14.301\pm0.046$ \\
         			K & $14.208\pm0.067$ \\
         			Y & $14.181\pm0.010$ \\
         			J$_\mathrm{s}$ & $14.251\pm0.013$\\
         			K$_\mathrm{s}$ & $14.309\pm0.008$\\
         			\hline
         		\end{tabular}
         		\label{tab:photometry}
         	\end{table}   
        
    \section{Spectral modeling with the PoWR stellar atmosphere code }
    \label{app:PoWR}
    
        PoWR assumes that the atmosphere, including the stellar wind, is stationary, spherically symmetric, energy conserving, and in radiative equilibrium. Hot star atmospheres cannot be approximated by LTE. Instead, the equations of radiative transfer and the statistical equilibrium have to be solved consistently in the comoving frame. This is achieved by iteration with the ``approximate lambda operator'' technique, yielding population numbers of the individual elements.
 	    
        The stellar atmosphere models which we apply here account for detailed model atoms of the elements: H, He, C, N, O, Mg, Si, P, S. The iron group elements Fe, Sc, Ti, V, Cr, Mn,  Co, and Ni are treated as one generic element (``G'') using a superlevel approach \citep[see][]{gra1:02}. The individual abundances for each element and the considered ionization levels adopted here are listed in Table~\ref{tab:elements}. 
 	    
 	    To calculate the detailed synthetic spectrum we assume a depth dependent micro-turbulence velocity. In our model calculations it is assumed that the micro-turbulence velocity starts with ${\zeta_\mathrm{ph}=\SI{10}{km\,s^{-1}}}$ in the photosphere and grows linearly with the wind velocity as ${0.1\,\varv(r)}$.
      
        The density stratification in the photosphere is calculated from the hydrostatic equations, taking the radiation pressure fully into account \citep{san2:15}. The velocity field in the supersonic regions, is adopted as a $\beta$-law \citep{cas1:75} with the standard exponent of $\beta=0.8$ \citep{pau1:86}.
 	    
        The stellar parameters, namely effective temperature $T_\mathrm{eff}$, surface gravity $g$, wind mass-loss rate $\dot{M}$, terminal wind velocity $\varv_\infty$, as well as the chemical composition, are determined by a comparison of the synthetic spectrum to the observation. The effective temperature $T_\mathrm{eff}$ is defined as referring to the radius where the Rosseland mean continuum optical depth is $\tau=2/3$. 
 	    
        Stellar winds are inhomogeneous; this is taken into account using the approximation for optically thin clumps (microclumping). The clumping factor $D$ describes by how much the density within the clumps is increased compared to a smooth wind with the same mass-loss rate \citep{ham2:98}. For our analysis, we adopt depth-dependent clumping  that starts at the sonic radius and increases outward, until a clumping factor of $D=10$ is reached at a radius of $10\,R_*$.
 	   
        \begin{table}[tbp]
         	\centering
         	\caption{Chemical elements and the ionization stages included in the PoWR models for AzV\,14.}
         	\begin{tabular}{cccc}\hline \hline \rule{0cm}{2.2ex}%
         		Element & Abundance & Ions & Reference\\
         		& [mass frac.]&&\\
         		\hline \rule{0cm}{2.8ex}%
         		\rule{0cm}{2.4ex}H  & $0.7375$           & \sc{i}, \sc{ii} & \citet{asp1:05}\\
         		\rule{0cm}{2.4ex}He & $0.2605$           & \sc{i}, \sc{ii}, \sc{iii} & $Y=1-X-Z$\\
         		\rule{0cm}{2.4ex}C  & $\num{21e-5}$      & \sc{iii} - \sc{vi} & \citet{tru1:07}\\
         		\rule{0cm}{2.4ex}N  & $\num{3e-5}$       & \sc{iii} - \sc{vi} & \citet{tru1:07}\\
         		\rule{0cm}{2.4ex}O  & $\num{113e-5}$     & \sc{ii} - \sc{vii} & \citet{tru1:07}\\
         		\rule{0cm}{2.4ex}Mg & $\num{10e-5}$      & \sc{ii} - \sc{v} & \citet{tru1:07}\\
         		\rule{0cm}{2.4ex}Si & $\num{13e-5}$      & \sc{iii} - \sc{vii} & \citet{tru1:07}\\
         		\rule{0cm}{2.4ex}P  & $\num{8e-7}$       & \sc{iv}, \sc{v}, \sc{vi} & \citet{sco2:15}$^{(a)}$\\
         		\rule{0cm}{2.4ex}S  & $\num{4e-5}$       & \sc{iv} - \sc{vii} & \citet{sco2:15}$^{(a)}$\\
         		\rule{0cm}{2.4ex}G  & $\num{35e-5}$      & \sc{iii} - \sc{ix} & \citet{tru1:07}\\
         		\rule{0cm}{2.0ex}   &                    &                    & \citet{sco2:15}$^{(a)}$\vspace{0.5ex}\\
         		\hline
         	\end{tabular}
         	\label{tab:elements}
            \begin{minipage}{0.95\linewidth}
                \ignorespaces 
                $^{(a)}$ We scaled their solar values by a factor of $1/7$ to account for the lower metallicity of the SMC.
            \end{minipage}
         \end{table}  

        \rule{0cm}{2cm}
    \section{Extraction of the TESS light curve}
    \label{sec:tess}
    
            AzV\,14 was observed with TESS, but has no automated pipeline processed light curve yet. Hence, we followed the method of \citet{bow1:22} and extracted pixel cutouts of size $20\times20$ using {\sc astrocut} \citep{Brasseur2019a} provided by MAST\footnote{https://archive.stsci.edu/}
            to extract a custom light curve and search for photometric variability. We tried differently sized aperture masks for an optimum light curve using the {\sc lightcurve} software package \citep{lig1:18}. Since AzV\,14 is a relatively faint target (${\mathrm{V}=\SI{13.9}{mag}}$) for TESS and because the pixel size of TESS is $\SI{21}{\arcsec}$, the aperture mask that maximized the signal-to-noise of the variability in the light curve and minimized the contribution of contaminating stars was found to be only a single pixel. We note that the field around AzV\,14 is quite crowded and that the employed TESS pixel contains several other stars. However, all stars identified by Gaia within this single neighboring pixels have $V\gtrsim16.5$. This is three orders of magnitude fainter than our target. None of this stars is known to be variable and combined they contribute less than $<5\%$ of the total measured flux. Pixel masks larger than a single pixel had larger percentage contamination and diminished signal from our target, thus justifying our extraction method. The background flux was estimated by calculating the median observed flux per frame whilst excluding pixels that predominantly contain stellar flux. The background flux was subtracted, and then we normalized the extracted light curve by dividing through the median flux. 
            
            TESS sector 1 data are known to be quite noisy and contain problematic systematics. For such a faint target this presented a significant challenge in extracting a reliable light curve. However, the light curves from sectors 27 and 28 are judged to be robust, and span a combined total of $\SI{50.8}{d}$. 
            
            The variability in the combined sectors 27 and 28 of the TESS data for AzV\,14 is well described as periodic. To determine the dominant periodicity we identified the highest-amplitude peak in the amplitude spectrum of the light curve. Next, we optimized the frequency and amplitude and determined uncertainties for this dominant periodicity by fitting a sinusoid to the light curve using nonlinear least-squares methods with the {\sc period04} software tool \citep{Lenz2005}. We determined the dominant frequency to be $\SI{0.53969\pm0.00019}{d^{-1}}$, corresponding to a period of $\SI{1.85292\pm0.00065}{d}$. Yet, as demonstrated from our spectroscopic analysis (see Sect.~\ref{sec:compare_spectra}), AzV\,14 contains two similar massive stars. Hence, periodic sinusoidal variability in such a system implies grazing eclipses or ellipsoidal variability. In phase space, however, one expects a symmetric double wave for such variability, meaning that the true dominant frequency is $\SI{1.07938\pm0.00038}{d^{-1}}$, and thus the orbital period of the system is ${P_\mathrm{orb}=\SI{3.7058\pm0.0013}{d}}$. The phase-folded light curve on this inferred orbital period is shown in Fig.~\ref{fig:phoebe}. We also find this frequency to be significant in the TESS data, but it has a smaller amplitude, hence is not the dominant frequency. It is not uncommon for the (sub)harmonic of the true orbital frequency to have a larger amplitude in the amplitude spectra of (eclipsing) binary star or ellipsoidal variable light curves \citep[see e.g.,][]{IJspeert2021a, Southworth2022a}.
            
            TESS has a low spatial resolution from its large pixel sizes, which when combined with our small aperture mask results in a relatively large scatter in the light curve. Nonetheless, it is possible to use the TESS data to constrain the inclination of the binary. In turn, this allows us to compare inclination dependent parameters, such as the projected rotation velocities to predictions of binary evolutionary models.   
    
    \onecolumn
    \section{Additional figures}
    
     	\begin{figure*}[thpb]
     	    \centering
     	    \includegraphics[width=\textwidth]{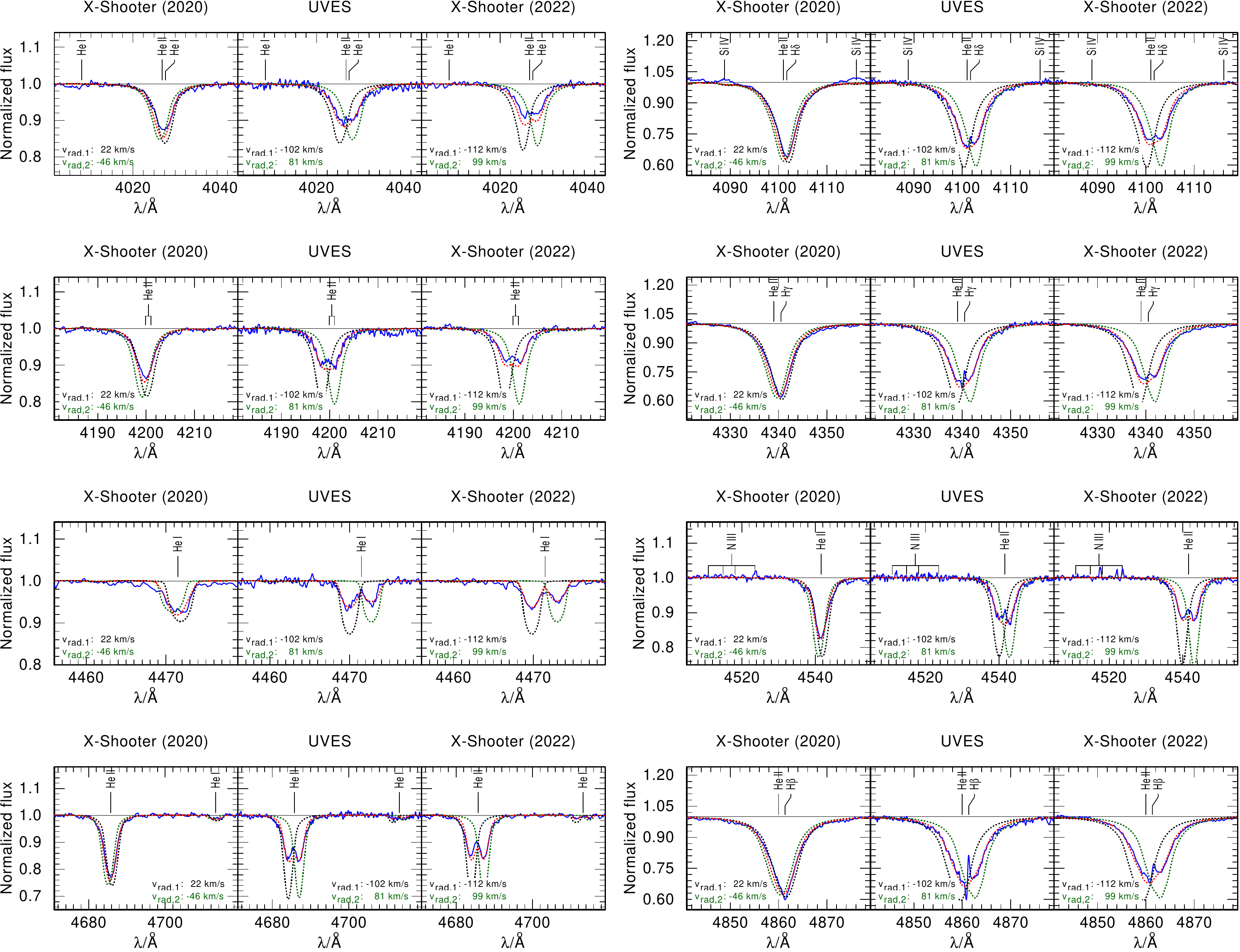}
     	    \caption{Hydrogen and helium lines in the two X-Shooter spectra and the UVES spectrum (blue) compared to the combined synthetic spectrum (red). The individual unweighted synthetic spectra of the primary and secondary components are represented as  black dotted and green dotted lines, respectively.}
     	    \label{fig:H_and_He_lines}
     	\end{figure*}
     	\begin{figure*}[ht!]
     	    \centering
     	    \includegraphics[width=0.95\textwidth]{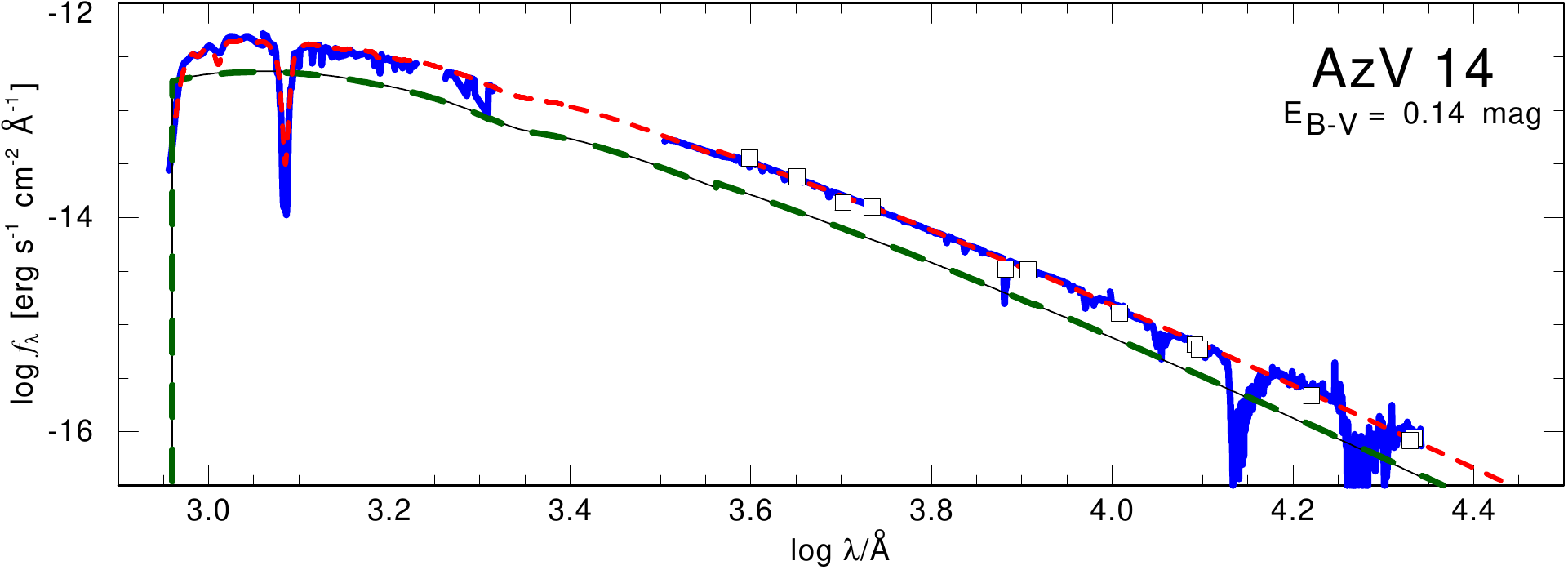}
     	    \caption{SED comparing flux-calibrated observations (blue lines) and photometric data (open squares) to  the model SED composed of both binary components (red dashed line). The individual synthetic SEDs of the primary and the secondary are shown as thin black and thick green dashed lines, respectively. Due to their similar luminosities these curves are hard to distinguish.}
     	    \label{fig:SED}
     	\end{figure*}
     	\begin{figure*}[thpb]
     	    \centering
     	    \includegraphics[trim= 0cm 0cm 0cm 0cm ,clip ,width=\textwidth]{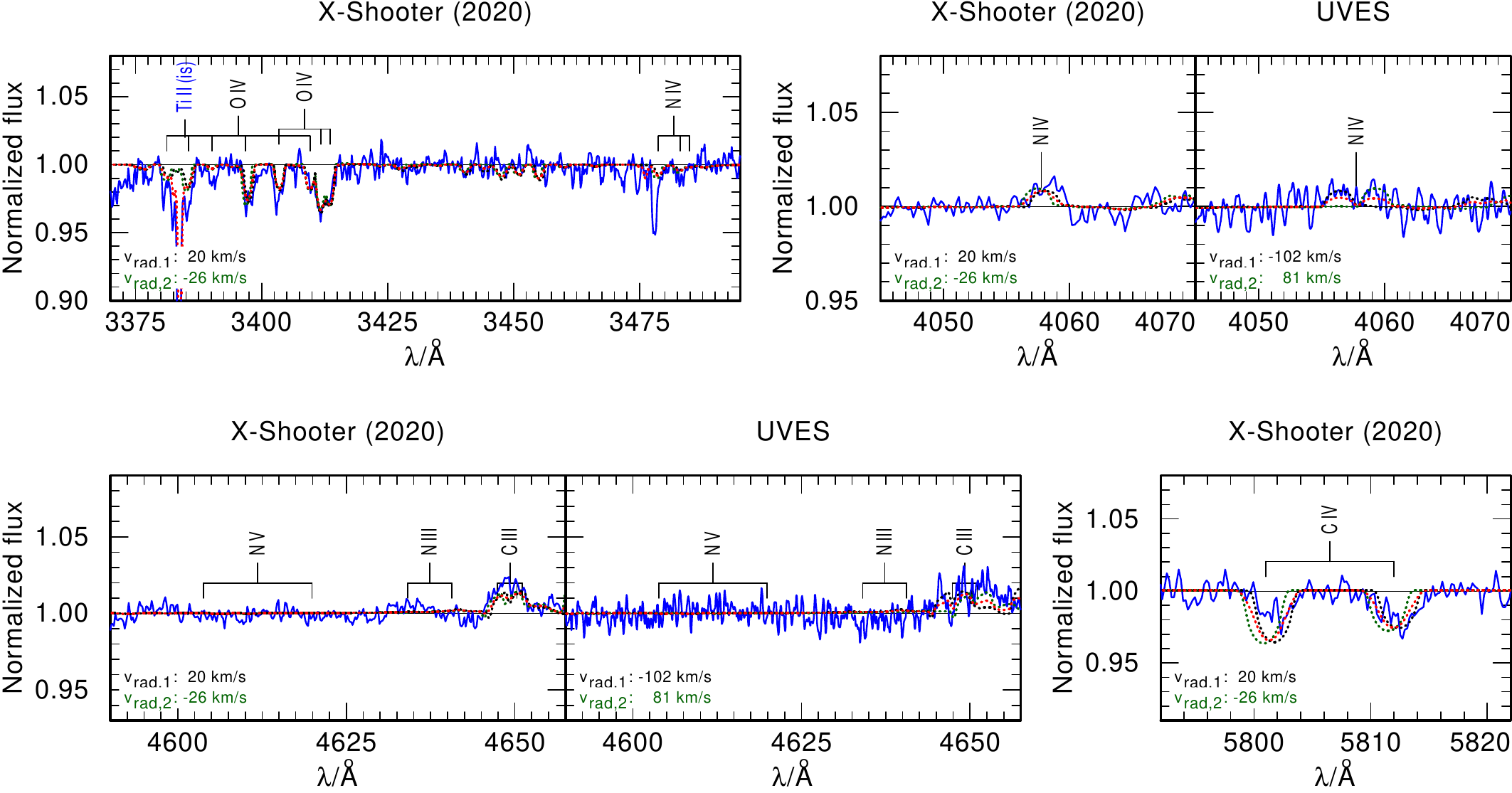}
     	    \caption{Close-up view of the metal lines in the X-Shooter~(2020) and UVES spectra. The observed spectrum is shown as a solid blue solid line, the individual unweighted synthetic spectra of the primary and the secondary component are represented as  black dotted and green dotted lines, respectively and the combined synthetic spectrum as red dashed line. In the combined synthetic spectrum ISM absorption lines, which originate from the Galactic foreground and in the SMC, are included. They are labeled as ``(is)''.}
     	    \label{fig:metal_lines}
     	\end{figure*}
     	\begin{figure*}[thpb]
     	    \centering
     	    \includegraphics[trim= -0.2cm 0cm 0cm -0.2cm ,clip ,width=\textwidth]{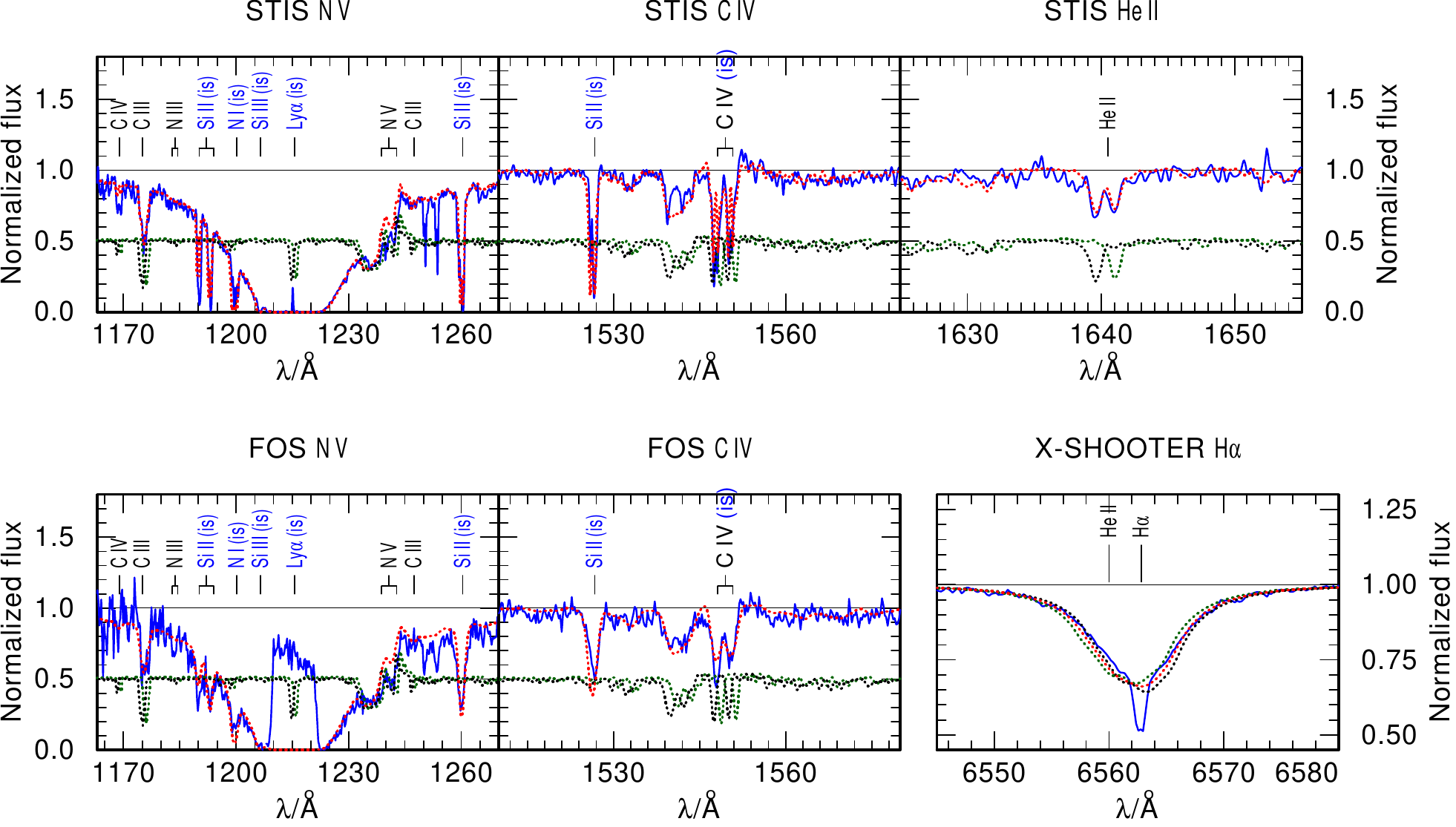}
     	    \caption{Close-up view of the wind diagnostic lines in the STIS and FOS spectra, as well as \Halpha{}, which is part of the X-Shooter (2020) spectrum. The observed spectrum is shown in blue and needs to be compared to the combined synthetic spectrum, shown in red. In the combined synthetic spectrum ISM absorption lines, which originate from the Galactic foreground and in the SMC, are included and labeled as ``(is)''. For the STIS and FOS spectrum we plot the weighted synthetic spectra. For the X-Shooter (2020) spectrum we show the  unweighted synthetic spectra of the primary and the secondary as black dotted and green dotted lines, respectively. A close-up view of the \HeII{}\,$\lambda4686$ wind diagnostic line can be found in Fig.~\ref{fig:H_and_He_lines}. }
     	    \label{fig:wind}
     	\end{figure*}
     	\begin{figure*}[thpb]
     	    \centering
     	    \includegraphics[trim= 0cm 0cm 0cm 0cm ,clip ,width=\textwidth]{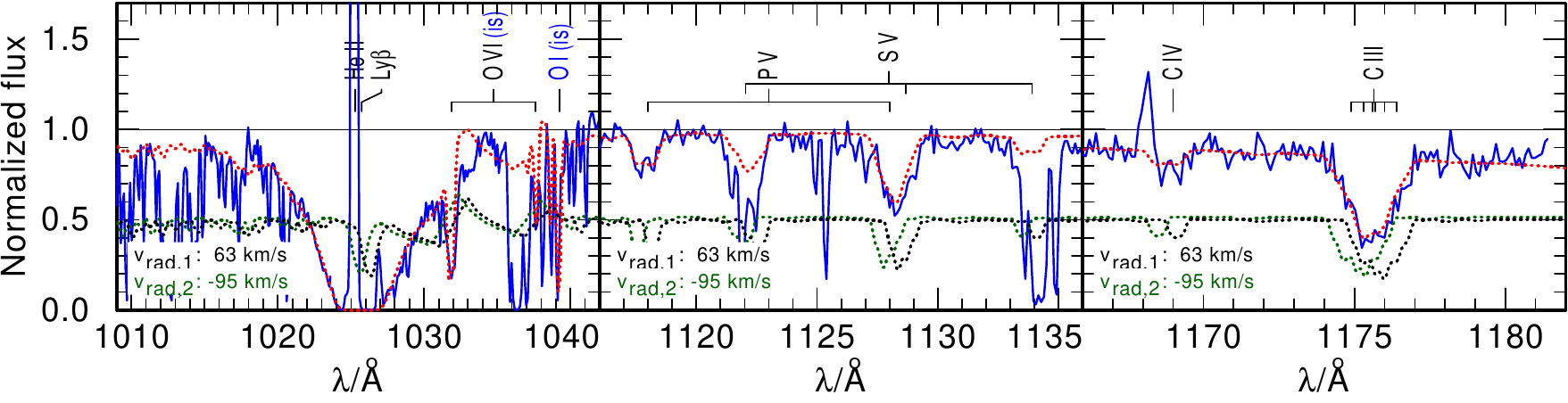}
     	    \caption{Close-up view of the important diagnostic lines in the far-UV FUSE spectrum. The observed spectrum is shown as a blue solid line, the weighted synthetic spectrum of the primary and secondary are shown as black dotted and green dotted lines and the combined synthetic spectrum is shown as a dashed red line. In the combined synthetic spectrum ISM absorption lines, which originate from the Galactic foreground and in the SMC, are included. They are labeled as ``(is)''. We note that not all ISM lines are modeled, and that FUSE spectra are additionally contaminated by absorption lines from molecular hydrogen.}
     	    \label{fig:fuse}
     	\end{figure*}
        \FloatBarrier
    \clearpage
    \twocolumn
    \section{Light and RV curve modeling with the PHOEBE code}
    \label{app:phoebe}
        
        The PHOEBE code  does not contain atmosphere tables for hot stars. Therefore, we approximated the AzV\,14 spectrum as a blackbody. From comparison of a blackbody and our synthetic spectrum in the V-band, we find that this is a reasonable approximation. To model the effect of limb-darkening, we adopt a quadratic approximation for the intensity \citep{dia1:92} 
        \begin{equation}
            I(\mu) = I(1)[1-a_i(1-\mu)-b_i(1-\mu)^2],
            \label{eq:ld}
        \end{equation}
        where $\mu=\cos\theta$ is the cosine of the directional angle and $a_i$ and $b_i$ are the linear and quadratic coefficients of the binary component $i$. The coefficients are determined by fitting the Eq.\,(\ref{eq:ld})  to the  emergent intensity distribution in our best fitting PoWR model. For the primary, we derive  ${a_1=0.1474}$ and ${b_1=0.0985}$,  and for the secondary, ${a_2=0.1489}$ and ${b_2=0.1025}$. The effect of gravitational darkening is approximated by a power-law with exponent $\beta_\mathrm{grav}=1$. For simplicity, we assume that both stars are rotating synchronously with the orbit, a reasonable approach for stars in binaries with orbital periods on the order of a few days. In our simulations, the effect of ellipsoidal variability, which is important in short-period binaries, is also included. The reflection effect \citep{wil1:90} is modeled with two reflections.

        As a fitting routine, we choose the built-in option of the \verb|emcee| sampler \citep{for1:13}, enabling us to consistently model the light and RV curve. From our PHOEBE model, we conclude that the observed light curve originates purely from ellipsoidal modulations, meaning that the flux during quadrature is the highest and lowest when the stars are aligned. The resulting parameters are listed in Table~\ref{tab:phoebe_results1} and \ref{tab:phoebe_results2}.

        \begin{table}[h]
            \footnotesize
        	\centering
        	\caption{Orbital parameters of AzV\,14 obtained using the PHOEBE code.}
        	\begin{tabular}{ccc}
        		\hline\hline \rule{0cm}{2.8ex}
        		\rule{0cm}{2.8ex}parameter & value & unit\\ 
        		\hline
        		\rule{0cm}{2.8ex}$P$ & $3.7058\pm0.0013$ (fixed)$^{(a)}$ & $[\mathrm{d}]$\\ 
        		\rule{0cm}{2.6ex}$\omega_0$ & $0\pm2$ & $[\mathrm{^\circ}]$\\ 
        		\rule{0cm}{2.6ex}$T_0$ & $\num{2459036.101\pm0.004}$ & $[\mathrm{d}]$\\ 
        		\rule{0cm}{2.6ex}$\varv_0$ & $-13\pm3$ & $[\mathrm{km\,s^{-1}}]$\\ 
        		\rule{0cm}{2.6ex}$q_\mathrm{orb}$ & $0.95\pm0.01$ &\\ 
        		\rule{0cm}{2.8ex}$i$ & $35^{+5}_{-5}$ & $[\mathrm{^\circ}]$\\ 
        		\rule{0cm}{2.6ex}$a$ & $40.6\pm1.8$ & $[{R_\odot}]$\vspace{1ex}\\ 
        		\hline 
        	\end{tabular}
     		\rule{0cm}{2.8ex}%
     		\begin{minipage}{0.95\linewidth}
         		\ignorespaces %
                    $^{(a)}$ Adopted from the analysis of the TESS light curve.
     		\end{minipage}
        	\label{tab:phoebe_results1}
        \end{table}  
        
        \begin{table}[h]
        \footnotesize
        	\centering
        	\caption{Orbital parameters of the individual binary components obtained using the PHOEBE code.}
        	\begin{tabular}{lccc}
        		\hline\hline \rule{0cm}{2.4ex}
        		\rule{0cm}{2.8ex}parameter & primary & secondary & unit\\ 
        		\hline
        		\rule{0cm}{2.8ex}$\quad\quad  K$ & $125\pm5$ & $132\pm5$ & $[\mathrm{km\,s^{-1}}]$\\ 
        		\rule{0cm}{2.6ex}$\quad\quad  M_\mathrm{orb}$ & $33.6^{+5.0}_{-3.7}$ & $31.9^{+4.8}_{-3.5}$& $[{M_\odot}]$\vspace{1ex}\\ 
        		\hline 
        	\end{tabular}
        	\rule{0cm}{2.8ex}%
     		\begin{minipage}{0.95\linewidth}
         		\ignorespaces
     		\end{minipage}
        	\label{tab:phoebe_results2}
        \end{table}
    \rule{0cm}{2cm}
 	\section{Input physics in the MESA binary evolutionary model}
 	\label{sec:input_physics}
 	
        Our binary evolutionary models are calculated with the MESA version r15140. The chemical composition is adjusted to the SMC. Following \citet{bro1:11}, we use tailored abundances for H, He, C, N, O, Mg, Si, and Fe in the evolutionary models. We note that these are similar to those used in the PoWR  atmosphere models (see Table~\ref{tab:elements}). As initial parameters, we only  set the initial masses of the binary components as well as the initial orbital period. We assume that the binary is initially tidally synchronized, a simplification adopted to reduce the free parameter space.
        
        Our models include rotational mixing and treat it as a diffusive process, which encompasses the effects of dynamical and secular shear instabilities, the Goldreich-Schubert-Fricke instability, as well as Eddingtion-Sweet circulations \citep{Heg1:00}. The efficiency coefficients of rotational mixing are adopted from \cite{bro1:11} and set to ${f_c = 1/30}$ and ${f_\mu = 0.1}$. On top of that, we include the angular momentum transport via magnetic fields from a Taylor-Spruit dynamo \citep{spr1:02}.
    
        We use the standard mixing length theory \citep{boe1:58} with a mixing length coefficient $\alpha_\mathrm{mlt}=1.5$ and the Ledoux criterion to model convection. Our models include the effect of envelope inflation, which occurs in regions within the star that approach the Eddington limit \citep{san1:15}. Semiconvective mixing is taken into account with an efficiency coefficient of $\alpha_\mathrm{sc}=1$ \citep{lan1:83,sch1:19} and for core-H burning we include step overshooting in such a way that the convective core is extended by $0.335H_\mathrm{P}$ \citep{bro1:11,sch1:19}. Thermohaline mixing, which becomes important for late evolutionary stages and during the accretion phase of the companion star, is modeled with an efficiency coefficient of $\alpha_\mathrm{th}=1$ \citep{kip1:80}. 
        
        The choice of mass-loss rates is crucial for stellar evolutionary models. In particular, the mass-loss rates  during the WR stages are important for those binaries which avoid evolving toward the cool side of the HRD. In the literature a discussion is ongoing on the efficiency of the WR winds at low metallicity.  For example,  \citet{san1:20} argue that below some limiting metallicity the WR winds become inefficient in removing mass. This limiting metallicity is thought to be close to the metallicity of the SMC. Therefore, we  adjust the WR mass-loss recipes such that they are adequate for the SMC metallicity. We include mass-loss as follows: for main sequence stars ($X_\mathrm{H}>0.7$) we use the mass-loss rates by \citet{vin1:01} (which are an order of magnitude higher than those empirically derived for AzV\,14 in this work). For temperatures below the so-called bi-stability jump we use the maximum value of either \citet{vin1:01} or \citet{nie1:90}. And for the WR phase, this is when the surface hydrogen abundance drops below $X_\mathrm{H}<0.4$, the recipe of \citet{she1:19,she2:19} is used. For surface abundances in the range of $0.7>X_\mathrm{H}>0.4$ we linearly interpolate between \citet{vin1:01} and \citet{she1:19,she2:19}.
        
        Mass transfer is modeled implicitly using the ``contact'' scheme of the MESA code. Mass transfer is treated as rotation-dependent mass accretion. This means, in our models mass transfer is conservative as long as the accretor can stay below critical rotation. As soon as it is rotating critically, we assume that all mass is lost from the system. In addition to that, we assume mass accretion from a stellar wind.

        The model calculation of a binary component is stopped after core-He depletion.
        It is assumed that as soon as the primary model depletes helium in its core, it directly collapses into a BH \citep{des1:11}. This means that the orbit is not perturbed by a supernova explosion. Under the assumption that we are able to model mass-transfer from the former secondary to the BH produced by the former primary. Recently it was reported that the first supernova originating from a WR progenitor was observed \citep{gal1:22}. We estimate that in the case of a supernova event the mass of the BH in our models might be overestimated by a factor 2 to 4. However, the main parameter defining the size of the Roche lobe, and thus the start and end of mass transfer is the orbital separation and not the mass ratio \citep[][their equation 2]{egg:83}. We assume that the impact of this simplification on the evolution of the former secondary is negligible. During the mass-transfer phase, we assume that the BH can only accrete material with the Eddington accretion rate.

    \section{Identifying WR phases in stellar evolutionary models}
    \label{app:WR_phase}
    
        \begin{table*}[t]
        \small
            \centering
            \caption{Summary of the stellar parameters of the WR stars in the SMC that are used to calibrate the threshold of the optical depth $\tau$.}
            \begin{tabular}{cccccccccccc}\hline \hline \rule{0cm}{2.8ex}%
                \rule{0cm}{2.2ex} star id & $T_\mathrm{eff}$ & $R_*$ & $\log L$ & $\log \dot{M}$ & $\log \dot{M}_\mathrm{Shenar}^{(a)}$ & $\varv_\infty$ & $X_\mathrm{H}$ & $\tau$ & binary &  reference \\
                \rule{0cm}{2.2ex}      [AB]   & $[\si{kK}]$ & $[\rsun]$ & $[\lsun]$ & $[\msunpyr]$ & $[\msunpyr]$ & $[\si{km\,s^{-1}}]$ & [mass fr.] & & & \\
                \hline \rule{0cm}{3.4ex}%
                \rule{0cm}{2.8ex}1   & 79$^{+10}_{-10}\,^{(b)}$ & 5.7$^{+1.3}_{-1.3}\,^{(c)}$ & $6.07\pm0.2$ & $-5.58\pm0.2$ & -5.33 & 1700$\pm200$ & $0.50\pm0.05$ & 0.26 & --- & \citet{hai1:15} \\
                \rule{0cm}{2.8ex}2   & $47^{+3}_{-3}$ & 9.1$^{+1.1}_{-1.1}\,^{(c)}$ & $5.57\pm0.1$ & $-5.75\pm0.2$ & -5.64 & 900$\pm100$  & $0.55\pm0.05$ & 0.19 & --- & \citet{hai1:15} \\
                \rule{0cm}{2.8ex}3   & $77^{+5}_{-5}$ & $5.0^{+1}_{-1}$ & $5.93\pm0.05$ & $-5.30\pm0.1$ & -5.23 & 1500$\pm200$ & $0.25\pm0.05$ & 0.53 &  x  & \citet{she1:16} \\
                \rule{0cm}{2.8ex}4   & $44^{+3}_{-3}$ & 13.0$^{+1.8}_{-1.8}\,^{(c)}$& $5.78\pm0.1$ & $-5.18\pm0.2$ & -5.14 & 1000$\pm100$ & $0.25\pm0.05$ & 0.37 & --- & \citet{hai1:15} \\
                \rule{0cm}{2.8ex}5a  & $43^{+5}_{-5}$ & $24.0^{+10}_{-7}$& $6.35\pm0.1$ & $-4.50\pm0.1$ & -4.80 & 2200$\pm200$ & $0.25\pm0.05$ & 0.51 &  x  & \citet{she1:16} \\
                \rule{0cm}{2.8ex}5b  & $43^{+10}_{-10}$ & $22.0^{+15}_{-10}$& $6.25\pm0.15$ & $-4.50\pm0.3$ & -4.86 & 2000$\pm200$ & $0.25\pm0.05$ & 0.60 &  x  & \citet{she1:16} \\
                \rule{0cm}{2.8ex}6   & $78^{+20}_{-5}$ & $4.7^{+1.5}_{-2.3}$ & $5.87\pm0.15$ & $-5.20\pm0.2$ & -5.27 & 2000$\pm200$ & $0.25\pm0.05$ & 0.56 &  x  & \citet{she1:18} \\
                \rule{0cm}{2.8ex}7   & $98^{+20}_{-10}$& $3.4^{+1.2}_{-1.2}$ & $6.10\pm0.1$ & $-5.00\pm0.2$ & -5.28 & 1700$\pm200$ & $0.15\pm0.05$ & 1.29 &  x  & \citet{she1:16} \\
                \rule{0cm}{2.8ex}9   & 99$^{+10}_{-10}\,^{(b)}$& 3.5$^{+0.8}_{-0.8}\,^{(c)}$ & $6.05\pm0.2$ & $-5.65\pm0.2$ & -5.23 & 1800$\pm200$ & $0.35\pm0.05$ & 0.31 & --- & \citet{hai1:15} \\
                \rule{0cm}{2.8ex}10  & 98$^{+10}_{-10}\,^{(b)}$& 2.2$^{+0.5}_{-0.5}\,^{(c)}$ & $5.65\pm0.2$ & $-5.85\pm0.2$ & -5.47 & 2000$\pm200$ & $0.35\pm0.05$ & 0.29 & --- & \citet{hai1:15} \\
                \rule{0cm}{2.8ex}11  & 89$^{+10}_{-10}\,^{(b)}$ & 3.5$^{+0.8}_{-0.8}\,^{(c)}$ & $5.85\pm0.2$ & $-5.56\pm0.2$ & -5.52 & 2200$\pm200$ & $0.40\pm0.05$ & 0.34 & --- & \citet{hai1:15} \\
                \rule{0cm}{2.8ex}12  & 112$^{+10}_{-10}\,^{(b)}$& 2.4$^{+0.5}_{-0.5}\,^{(c)}$ & $5.90\pm0.2$ & $-5.79\pm0.2$ & -5.45 & 1800$\pm200$ & $0.20\pm0.05$ & 0.30 & --- & \citet{hai1:15}\vspace{0.75ex}\\
                \hline
            \end{tabular}
            \rule{0cm}{2.8ex}%
            \begin{minipage}{0.95\linewidth}
                \ignorespaces 
                $^{(a)}$ Calculated using the mass-loss recipe of \citet{she1:19,she2:19}. $^{(b)}$ We adopt a typical uncertainty of $\SI{10}{kK}$ (i.e., this is equivalent to one step in their used model grid). $^{(c)}$ Calculated using Gaussian error propagation.
            \end{minipage}
            \label{tab:WR_stars}
        \end{table*}
 	
        We calculated the optical depth at the surface of a star in our evolutionary model calculation with the formula from \citet{lan1:89}
 	    \begin{equation}
 	        \tau = \dfrac{-\kappa\dot{M}}{4\pi R(\varv_\infty-\varv_0)}\ln\left(\dfrac{\varv_\infty}{\varv_0}\right).
 	    \end{equation}
        In this equation, the opacity $\kappa$ is approximated by the electron scattering opacity ${\kappa_\mathrm{es}=0.2\,(1+X_\mathrm{H})\,\mathrm{cm^2\,g^{-1}}}$, $\dot{M}$ is the wind mass-loss rate, $R$ is the radius of the star, $\varv_\infty$ is the terminal wind velocity, and $\varv_0$ is the expansion velocity at the base of the wind, which, for simplicity, is assumed to be constant ${\varv_0=\SI{20}{km\,s^{-1}}}$. \citet{gra1:17} found that $\varv_\infty$ can be linked to the escape velocity $\varv_\mathrm{esc}$ by
 	    \begin{equation}
 	        \varv_\infty=\alpha\cdot\varv_\mathrm{esc} = \alpha\cdot\sqrt{\dfrac{2GM}{R}(1-\Gamma)},
 	        \label{eq:vinf}
 	    \end{equation}
        where $\Gamma$ is the Eddington factor and $\alpha$ is an evolutionary dependent scaling factor. We assume $\alpha=1.3$ during the main sequence and WN phase. For factors used in the other phases we refer to \citet{pau1:22}, their section 2.2.
 	    
        Ideally the threshold to differentiate between optically thin and thick winds, He-stars or WR stars, is $\tau\geq1$. However, in the calculation of the optical depth several simplifications are included. Therefore, a calibration to observations is needed. \citet{agu1:22} already did this for H-free WN stars and determined a threshold value of $\tau\geq1.45$. On the other side, \citet{pau1:22} find that for H-poor WN stars a lower threshold might be needed to reproduce observations. In the SMC only H-poor WN stars (and one WO type star) are present. To find a threshold value that can reproduce the observations, we calculated the optical depths of all the H-poor WN stars in the SMC. Their stellar parameters and optical depths are listed in Table~\ref{tab:WR_stars}. We find that all have an optical depth value of $\tau\gtrsim0.2$, which we set here as our new threshold.

    \end{appendix}

\end{document}